\documentclass[conference]{IEEEtran}
\IEEEoverridecommandlockouts
\usepackage{color}
\usepackage{graphicx}  
\usepackage{psfrag}    
\usepackage{algorithm}
\usepackage{algorithmic}
\usepackage{amsfonts,amssymb}
\usepackage{amsthm}
\usepackage{amsmath}
\usepackage{amstext}
\usepackage{bm}
\usepackage{subfigure}
\usepackage{hyperref}
\usepackage{CJK}
\usepackage{url}
\usepackage{lipsum}
\usepackage{cuted}
\usepackage{cite}
\usepackage{cases}

\begin{document}

\title{Joint Spectrum and Power Allocation for V2X Communications with Imperfect CSI}
\author{Peng Wang, Weihua Wu, Jiayi Liu, Guanhua Chai and Li Feng
\thanks{*This work was supported in part by the NSF China under Grant 61801365, in part by the China Postdoctoral Science Foundation under Grant 2018M643581 and in part by Jiangsu Province Postdoctoral Foundation (No. 2020Z208).}
\thanks{P. Wang, W. Wu, J. Liu and G. Chai are with State Key Laboratory of ISN, School of Telecommunications Engineering, Xidian University, No.2 South Taibai Road, Xi'an, 710071, Shaanxi, China. (Email:whwu@xidian.edu.cn).}
\thanks{L. Feng is with School of Computer Science and Communication Engineering, Jiangsu University, No. 301 Xuefulu, Zhenjiang, 212013, Jiangsu, China.}

}
\maketitle


\begin{abstract}
In Vehicle-to-Everything (V2X) communication, the high mobility of vehicles generates the Doppler shift which leads to channel uncertainties. Moreover, the reasons for channel uncertainties also include the finite channel feedback, channels state information (CSI) loss and latency.
With this concern, we formulate a joint spectrum and power allocation problem for V2X communication with imperfect CSI. Specifically, the sum capacity of cellular user equipments (CUEs) is maximized subject to the minimum  Signal-to-Interference-and-Noise Ratio (SINR) requirements of CUEs and the outage probability constraints of vehicular user equipments (VUEs). Then, two different robust resource allocation approaches are designed to solve the problem. One is Bernstein Approximation-based Robust Resource Allocation approach. More specifically, Bernstein approximations are employed to convert the chance constraint into a calculable constraint, and Bisection search method is proposed to obtain the optimal allocation solution with low complexity.
Then, for further reducing the computational complexity, Self-learning Robust Resource Allocation approach, which includes a learning method and an analytical mapping method, is proposed as the second approach. The learning method is devised to learn the uncertainty set which transforms the chance constraint into calculable constraints, and the analytical mapping method is proposed to obtain closed-form solutions of the resource allocation problem. Finally, the simulation results prove that the proposed approaches can improve the capacity of all CUEs effectively whilst ensuring the reliability of the channel.


\vspace{5pt}
\textbf{\emph{Key Terms}}: resource allocation, imperfect CSI, V2X communications, chance constraint, robust optimization.

\end{abstract}


%
\section{Introduction}

\newtheorem {theorem}{\textbf{Theorem}}
\newtheorem {lemma}{\textbf{Lemma}}
\newtheorem {remark}{\textbf{Remark}}
\newtheorem {definition}{\textbf{Definition}}
In recent years, with the evolution of the significant research area about vehicular networks, automotive and telecommunication industries have begun to focus on this research.
The application of wireless communication to vehicular networks has generated a new concept denoted as Vehicle-to-Everything (V2X) communication. V2X communication consists of Vehicle-to-Infrastructure (V2I), Vehicle-to-Vehicle (V2V), Vehicle-to-Pedestrian (V2P) and so on\cite{9063405,8731910}.
The goal of V2X communication is to assist traffic management and guarantee road security. In addition, it can be applied to autopilot and vehicular entertainment. V2X communication demands to achieve the communication in real-time between vehicles and other objects, such as other vehicles, base stations (BS) and intelligent mobiles. As a result, it is considered as a key technique to connect vehicles and satisfy the requirement of intelligent transportation system (ITS) \cite{887fdfdf4,SHARMA2019100182}.

To satisfy the requirements of V2X communication, two leading wireless communication technologies are put forward by the research community. The WiFi technology, which works in 5.9 GHz, is supported by the IEEE 802.11p standard \cite{sehla2022resource}.
Vehicular Ad-hoc NETworks, which apply the WiFi technology as the first standard, can enable vehicles to access the network in a competitive way.
The cellular technology, which works in permissible bands of networks, is represented as Cellular-V2X (C-V2X) \cite{etsi2018103}.
The Third Generation Partnership Program (3GPP) proposed the Long Term Evolution V2X (LTE-V2X) where radio resources were managed uniformly by the BS in 2016\cite{111}.
Afterwards, the 3GPP standardized the 5G New Radio-V2X (NR-V2X) in Release 16 in 2018 \cite{8998153}. Compared with 802.11p, the analysis shows that C-V2X has various advantages in the field of latency, coverage area and throughput \cite{machardy2018v2x,8490729}. Thus, NR-V2X is chosen as the communication technology in this work.

In practice, it is inevitable to face the challenge of the resource allocation problem when V2X communication is applied.
Hence, the research community has devoted in the V2X resource allocation problem in recent years.
Generally, the radio frequency is always scarce and essential so that it should be efficiently allocated.
Since the BS cannot cover all areas, NR-V2X proposes two modes to allocate resources: the centralized under-coverage mode (mode-1) and the distributed out-of-coverage mode (mode-2) \cite{garcia2021tutorial}.
Based on these two modes, vehicles can directly communicate with each other bypassing BS.
The difference between the two modes is their arrangement approaches about the radio frequency. Specifically, the Next Generation NodeB (gNB) in mode-1 is in charge of scheduling and allocating resources to vehicles.
On the contrary, the vehicles in mode-2 can autonomously choose their own radio resources.
Compared with mode-2, mode-1 can efficiently use sub-carriers with less interference because the gNB is able to collect comprehensive information from the vehicles covered by the gNB. Therefore, this paper studies the resource allocation problem based on mode-1 of NR-V2X networks.

The resource arrangement for V2X communication faces distinct challenges compared with traditional wireless communication. In practical situations, an important feature of V2X communication is the high mobility of vehicles which leads to the Doppler effect in small-scale fading.
Therefore, acquiring the precise channel state information (CSI) is challenging. Particularly, when the links are not directly connected to the gNB, the precise CSI is more difficult to be acquired due to latency.
Typically, for vehicles running at low speeds, the Doppler shift is not prominent and is neglected. Nevertheless, for vehicles running at high speeds, it can significantly affect the small-scale fading which results in rapid channel variations \cite{8638940}.
Moreover, the challenge of acquiring the precise CSI should also include the finite channel feedback, CSI loss and latency \cite{mo2018limited,pourkabirian2021robust,lin2021capacity}.
To sum up, the channel uncertainty can result in the violation of V2V links and effect the resource arrangement of V2X communication.

Our work investigates the resource allocation of V2X communication with imperfect CSI.
In realistic vehicular networks, the gNB can merely acquire the precise CSI in V2I links, whereas it estimates the CSI in V2V links with some errors due to the uncertainty of CSI.
To overcome the problem of the limited spectrum, we present a system model where each cellular user equipment (CUE) can share the spectrum with at most one vehicular user equipment (VUE), and each VUE can reuse the spectrum of at most one CUE. Then, we design a joint spectrum and power allocation optimization problem to maximize the sum capacity of CUEs subject to the minimum Signal-to-Interference-and-Noise Ratio (SINR) requirement of CUEs and the probabilistic quality of service (QoS) of VUEs. The contributions of this work are summarized as follows:

\begin{itemize}
  \item Bernstein Approximation-based Robust Resource Allocation approach is presented to solve the resource allocation problem. Specifically, Bernstein approximations are employed to make the chance constraint tractable, and Bisection search method is developed to resolve the non-convex problem with low complexity.
  \item For further reducing the computational complexity, Self-learning Robust Resource Allocation approach, which includes a learning method and an analytical mapping method, is proposed to resolve the above problem. We devise the learning method, which learns the channel uncertainty set to obtain an affine set, to transform the intractable chance constraint into simple linear constraints. The analytical mapping method is developed to obtain closed-form solutions of the non-convex optimization problem.
  \item The trade-off between robustness and the capacity of CUEs is revealed for the above two approaches. Simulation results are provided to demonstrate that Self-learning Robust Resource Allocation approach increases robustness at the expense of decreasing the capacity of CUEs. On the contrary, Bernstein Approximation-based Robust Resource Allocation approach increases the capacity of CUEs at the expense of decreasing robustness.
\end{itemize}

The remainder of this paper is organized as follows. In Section II, we review the related work. We introduce the system model and state the optimization problem in Section III.
In section IV, Bernstein Approximation-based Robust Resource Allocation approach is presented to resolve the optimization problem.
For further minimizing the computation complexity, Self-learning Robust Resource Allocation approach is proposed in Section V.
Then, the suboptimality of the proposed methods is analysed in Sections VI.
Finally, we elaborate the simulation results in Sections VII and summarize the paper in Sections VIII.

\section{Related Work}

The resource allocation is a significant part of the challenges for V2X communications.
To effectively reduce the inter-channel interference and markedly enhance the throughput in V2X networks, the applicable resource allocation schemes should be adopted \cite{8891446}.
Therefore, for achieving the ultra-reliable V2X communications, different resource management approaches have been researched in \cite{8731910,9129179,8732472,9103244,chen2020fsfsf}.
In \cite{8731910}, the resource management scheme, which utilized the expressions of reliability and delay assurances, was presented to guarantee the QoS requirements for V2X communication.
For ensuring the SINR to exceed its objective threshold, the work in \cite{9129179} proposed a two-stage resource allocation method for V2X communication.
The paper in \cite{8732472} designed a mode selection method based on the resource management whilst the minimum SINR requirements and the maximum global information value were guaranteed for V2X communication.
The authors in \cite{9103244} developed a jointly power and subcarrier allocation approach to guarantee the minimum SINR requirements for C-V2X communication which is satisfied by the long-term SINR.
The work in \cite{chen2020fsfsf} studied the resource allocation for intelligent reflecting surface aided vehicular communications while the minimum SINR of V2V links was guaranteed.

Nevertheless, in above approaches, the minimum SINR requirements could merely supply the best-effort services for VUEs.
Thus, these approaches failed to adopt the form of probability to guarantee the requirements of the ultra-reliable transmission.
For guaranteing the VUE QoS requirements, the outage probability of the SINR requirements must be below the default probability.
In \cite{8664606}, Bernstein approximations were adopted to transform the probabilistic constraints of the interference of D2D-V links into the tractable constraints, and the resource allocation problem was resolved by the successive convex approximation.
The authors in \cite{chen2021fsdfss} developed the block coordinate descent approach to solve the spectrum sharing problem in RIS-aided vehicular networks where the outage probability of V2V links was employed to ensure the reliability of V2V communication.
The work in \cite{8345717} utilized the chance constraints to ensure the minimum reliability requirement for V2V and proposed a suite of graph-based resource management algorithms to resolve the resource allocation problem.
The reliability requirements for VUEs, expressed as the tolerable VUEs chance constraints, were converted into an easier affine constraint in \cite{7913583}, and the analytical mapping method was designed to obtain the optimal transmit power for V2X networks.

However, all of these works neglected the channel uncertainties, resulting in the violation of QoS constraints for V2V links. Though some researchers used the robust optimization to handle the CSI uncertainties in \cite{chen2021rfgdgst,li2020refdfdrce,peng2021roddfgdfst}, they considered the error region of CSI uncertainty as the prior knowledge and the CSI uncertain set as the fixed-size set. In a realistic scenario, due to the unknown of the channel uncertainty model, these approaches were challenging to be deployed in the fast variational channels for the V2X communication. Although the resource allocation based on the uncertain probability distribution information was solved in \cite{9382930,wu2021robust}, they required the mean and variance of the uncertain CSI samples, which complicated the resource allocation approaches and increased the computational complexity. Thus, this paper investigates the robust resource allocation problem based on the uncertain CSI in V2X networks with low complexity.

\section{System Model}

In this section, we introduce the system model and discuss the channel uncertainty scenarios in V2X communication network.
Then, the resource allocation problem is formulated based on the different QoS requirements.

\subsection{Network Model}

In Fig. 1, a single cellular vehicle network is considered as the system model. We assume there are $J$ vehicles whose set is $ {\cal J}= \{ 1, \cdots ,J\} $, denoted as CUEs. The CUEs apply V2I links to deliver messages with the gNB.
There are $S$  pairs of vehicles whose set is $ {\cal S}= \{ 1,\; \cdots ,S\} $, denoted as VUEs.
The VUEs achieve the V2V communication. Furthermore, the Uu interface of NR is utilized to send messages for V2I communication and the transmission mode-1 in NR sidelink is utilized for V2V communication.
The set of resource blocks (RBs) is allocated to Uu interface by the gNB.
Orthogonal multiple access technique is applied in the above network model so that each orthogonal RB can serve a vehicle under OFDMA subcarriers. The orthogonal RBs can weaken the interference of neighboring CUEs. Because the CUEs sparsely use the uplink resource and the VUEs can cause less interference for gNB, the VUEs can reuse the uplink resource of CUEs.
The indicator variable ${\rho_{j,s}}$ represents the allocation of the uplink resource for VUEs. Significantly, if the spectrum of the $j$th CUE is reused by the $s$th VUE then ${\rho_{j,s}}=1$, otherwise ${\rho_{j,s}} = 0$.
$p_j^c$ represents the transmit power of the $j$th CUE. Likewise, $p_s^d$ represents the transmit power of the transmitter of the $s$th VUE.
We represent the channel gain between the $s$th VUE as
\begin{figure}
\begin{center}
\includegraphics[scale = 0.3]{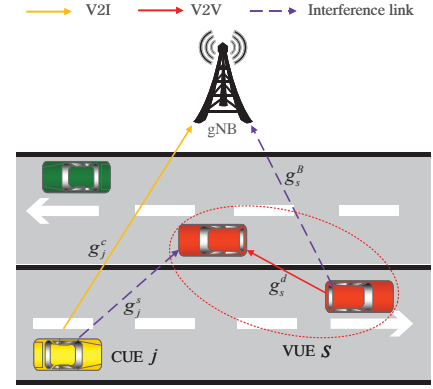}
\label{fig1}\caption{Single cellular vehicle network of V2X communication.}
\end{center}
\vspace{-2em}
\end{figure}
\begin{eqnarray}
g_s^d = {\left| {h_s^d} \right|^2}\varpi _s^d\kappa D_s^{ - \epsilon } \buildrel \Delta \over = {\left| {h_s^d} \right|^2}\omega _s^d,
\end{eqnarray}
where $\omega _s^d = \varpi _s^d\kappa D_s^{ - \epsilon }$ denotes the large-scale slow fading channel gain, $h_s^d$ represents the small-scale fast fading power component, $\kappa $ is the pathloss constant, $D_s$ is the distance between the $s$th VUE, $\epsilon $ is the decay exponent, and $\varpi _s^d$ is log-normal shade fading random variable with a standard deviation $\vartheta $. Similarly, $g_j^c$ denotes the channel gain between the $j$th CUE and the gNB, $g_j^s$ denotes the crosstalk channel gain between the $j$th CUE and the $s$th VUE, $g_s^B$ denotes the crosstalk channel gain between the $s$th VUE and the gNB and their expression forms are all similar to $g_s^d$.

Because the large-scale fading components generally rely on the vehicular position and undergo the slow scale variation, the gNB can entirely acquire the large-scale fading information of all communication links.
However, we make different assumptions for the small-scale fading of different links.
We assume that the small-scale fading information $h_j^c$ and $h_s^B$ of the links can be accurately obtained at the gNB in this paper. The reasons are mainly the following two fold. On the one hand, since the links $g_j^c$ and $g_s^B$ are directly connected to the gNB as shown in Fig. 1, their latency is very low. On the other hand, since the gNB is stationary in the side of the road, the relative movement between the gNB and the transmitters of the vehicles is marginal when compared with the relative movement between the transmitters and the receivers of the vehicles.
For acquiring the small-scale fading $h_s^d$ and $h_j^s$ of the links, we present the physical sidelink feedback channel (PSFCH), which is applied to transmit the CSI of the receiver to the transmitter.
When the PSFCH receives the reply, a physical uplink control channel (PUCCH), which forwards the sidelink CSI from transmitter to the gNB, is allocated to the transmitter by the gNB.
There is some delay when the CSI is transmitted to the gNB by vehicular links because of the overhead of establishment and disconnection of the channels.
Moreover, the high mobility of vehicles can result in the Doppler shift which affects the small-scale fading.
Therefore in this paper, we suppose that gNB merely acquires the estimated small-scale fading $\hat h_s^d$ and $\hat h_j^s$ with estimation error $e_s^d$ and $e_j^s$, respectively.
The additive error \cite{kim2010does} model of the channel imperfection is applied, i.e.,
\begin{eqnarray}
h = \lambda \hat h + \sqrt {1 - {\lambda ^2}} e,
\end{eqnarray}
where $e\sim {\cal C}{\cal N}(0,1)$ is i.i.d., $\lambda\ (0 < \lambda  < 1)$ expresses the channel estimation error coefficient. From the fading channel of the Jakes \cite{liu2021resource} statistical model, $\lambda $ is expressed as
\begin{eqnarray}
\lambda  = {J_0}(2\pi {f_s}T),
\end{eqnarray}
where ${J_0}$ is the zero-order Bessle function of the first kind and $T$ is the channel feedback latency. ${f_s} = v{f_v}/c$ represents the maximum Doppler frequency, where $v$ denotes the vehicular speed, $f_v$ is the carrier fequency and $c = 3 \times {10^8}{\rm{m/s}}$. This paper supposes that the transmitting vehicles and the receiving vehicles can obtain the accurate $\lambda $.

On this basis, the received SINR of the $s$th VUE is expressed as
\begin{eqnarray}
\Gamma_s^d = \frac{{p_s^d\omega _s^d\left( {{\lambda ^2}{{\left| {\hat h_s^d} \right|}^2} + (1 - {\lambda ^2}){{\left| {e_s^d} \right|}^2}} \right)}}{{{\sigma ^2} + \sum\limits_{j \in {\cal J}} {{\rho_{j,s}}p_j^c\omega _j^s\left( {{\lambda ^2}{{\left| {\hat h_j^s} \right|}^2} + (1 - {\lambda ^2}){{\left| {e_j^s} \right|}^2}} \right)} }},
\end{eqnarray}
where ${\sigma ^2}$ is the power of the additive white Gaussian noise. Likewise, the SINR from the $j$th CUE to the gNB is expressed as
\begin{eqnarray}
\Gamma_j^c = \frac{{p_j^c{{\left| {h_j^c} \right|}^2}\omega _j^c}}{{{\sigma ^2} + \sum\limits_{s \in {\cal S}} {{\rho_{j,s}}p_s^d{{\left| {h_s^B} \right|}^2}\omega _s^B} }}.
\end{eqnarray}

\subsection{Problem Formulation}

Because the V2I communications can satisfy the requirement of mobile high-data rate services, it is significant to maximize the capacity of all CUEs.
In V2X communication systems, the V2I communications usually serve insecure scenarios. The reliability of V2V communications is ensured by the probability of outage event since the CSI of vehicles is uncertain. Therefore, the resource allocation problem can be formulated as
\begin{subequations}
\begin{align}
\max\limits_{\{ {\rho_{j,s}}\} \{ p_j^c\} \{ p_s^d\} } &\sum\limits_{j = 1}^J B_c {\log _2}(1 + \Gamma _j^c)\\
\textrm{s.t.} \quad \ \ \
&\Gamma _j^c \ge \Gamma _{\min }^c,\forall j \in {\cal J},\\
&  \Pr \{ \Gamma _s^d \ge \Gamma _{\min }^d\}  \ge 1 - \beta ,\forall s \in {\cal S},\\
&  \sum\limits_{s = 1}^S {{\rho_{j,s}} \le 1,{\rho_{j,s}}}  \in \{ 0,1\} ,\forall j \in {\cal J},\\
&  \sum\limits_{j = 1}^J {{\rho_{j,s}} \le 1} ,\forall s \in {\cal S},\\
& 0 \le p_j^c \le p_{\max }^c,\forall j \in {\cal J},\\
& 0 \le p_s^d \le p_{\max }^d,\forall s \in {\cal S},
\end{align}
\end{subequations}
where $\Gamma _{\min }^c$ is the SINR threshold of V2I communications, $\Gamma _{\min }^d$ is the SINR threshold of V2V communications, $B_c$ denotes the channel bandwidth of RBs, $Pr\{\cdot\}$ denotes the form of the probability, $\beta $ represents the maximum acceptable outage probability of V2V communications, $p_{\max }^c$ and $p_{\max }^d$ denote the maximum transmit power of V2I communications and V2V communications, respectively.
Constraints (6d) and (6e) express that each CUE can share the spectrum with at most one VUE pair and each VUE pair can only reuse the spectrum of at most one CUE, respectively.

For solving problem (6), the major challenges are listed in following: (i) Obtaining the closed-form expression of (6c) is difficult due to the existence of the outage probability constraint and the channel uncertainty for V2V communications; (ii) Problem (6) is a non-convex problem since (6a) is non-convex and the variable $\rho_{j,s}$ belongs to the discrete variable; (iii) The variables in problem (6) are coupled with each other since the variables $\rho_{j,s}$, $p_j^c$ and $p_s^d$ coexist in the numerators and denominators of $\Gamma _j^c$ and $\Gamma _s^d$. To tackle these challenges, we propose two different robust resource allocation approaches to solve problem (6).

\section{Bernstein Approximation-based Robust Resource Allocation}

As mentioned in above sections, the accurate distributions of $e_s^d$ and $e_j^s$ are difficult to receive in practical vehicular communication. In addition, the different V2X scenarios can generate different distributions of $e_s^d$ and $e_j^s$. In this section, we suppose that the distributions of $e_s^d$ and $e_j^s$ are uncertain.
Then, problem (6), which can be decomposed into a power allocation problem and a spectrum allocation problem, is solved by Bernstein Approximation-based Robust Resource Allocation approach.
To be specific, the optimal power allocation solutions can be obtained based on each possible spectrum reusing pair of CUE and VUE.
Then, the optimal spectrum allocation solution can be obtained based on the optimal power allocation solutions of all possible reusing pairs.

\subsection{Chance Constraint Approach}

The power allocation problem is researched based on each possible reusing pair of CUE and VUE in following subsection. Assuming that the spectrum reusing pair is determined, the power allocation problem is simplified as
\begin{subequations}
\begin{align}
C_{j,s} = \max \limits_{\{ p_j^c\} \{ p_s^d\} } &B_c{\log _2}(1 + \frac{{p_j^cg_j^c}}{{{\sigma ^2} + p_s^dg_s^B}})\\
\textrm{s.t.}\quad
&   \frac{{p_j^cg_j^c}}{{{\sigma ^2} + p_s^dg_s^B}} \ge \Gamma _{\min }^c,\label{7b}\\
&  \Pr \{ \frac{{p_s^dg_s^d}}{{{\sigma ^2} + p_j^cg_j^s}} \ge \Gamma _{\min }^d\}  \ge 1 - \beta ,\label{7c}\\
& 0 \le p_j^c \le p_{\max }^c,0 \le p_s^d \le p_{\max }^d. \label{7d}
\end{align}
\end{subequations}
For better compatibility with the following robust optimization approaches, constraint (\ref{7b}) is translated into a linear constraint which can be rewritten as
\begin{eqnarray}
\frac{{p_j^cg_j^c}}{{{\sigma ^2} + p_s^dg_s^B}} \ge \Gamma _{\min }^c \Rightarrow \frac{{p_j^cg_j^c}}{{\Gamma _{\min }^c}}{\rm{ - }}p_s^dg_s^B \ge {\sigma ^2}. \label{8}
\end{eqnarray}
Likewise, the content in the brace of constraint (\ref{7c}) can be rewritten as
\begin{eqnarray}
\frac{{p_s^dg_s^d}}{{{\sigma ^2} + p_j^cg_j^s}} \ge \Gamma _{\min }^d \Rightarrow
\frac{{p_s^dg_s^d}}{{\Gamma _{\min }^d }} - p_j^cg_j^s \ge {\sigma ^2}
\Rightarrow
\mathbf{p}_s^d\boldsymbol{\theta} _s^d \ge {\sigma ^2},  \label{adsdfsf}
\end{eqnarray}
where $\mathbf{p}_s^d = [\frac{{p_s^d}}{{\Gamma _{\min }^d}}, - p_j^c]$ and $\boldsymbol{\theta} _s^d = {[g_s^d,g_j^s]^T}$.
As shown in \cite{article3}, a convex function can take the place of (\ref{7c}) as the safe approximations based on the Bernstein approach for i.i.d. uncertain channel parameters. The uncertainty set can be written as
\begin{eqnarray}
\boldsymbol{\theta} _s^d \in [ \boldsymbol{\bar\theta} _s^d - \boldsymbol{\hat \theta} _s^d,\boldsymbol{\bar \theta} _s^d + \boldsymbol{\hat \theta} _s^d],
\end{eqnarray}
where $\boldsymbol{\bar \theta} _s^d = {[\bar g_s^d,\bar g_j^s]^T}$, $\boldsymbol{\hat \theta} _s^d = {[\hat g_s^d,\hat g_j^s]^T}$, $\bar g_s^d$ and $\bar g_j^s$ are the estimated values of $g_s^d$ and $g_j^s$, respectively. $\hat g_s^d$ and $\hat g_j^s$ are the error values of $g_s^d$ and $g_j^s$, respectively. For applying the chance constraint approach, the V2V constraint in (\ref{7c}) is rewritten as
\begin{eqnarray}
\boldsymbol{p}_s^d\boldsymbol{\theta} _s^d = \frac{{p_s^d\bar g_s^d}}{{\Gamma _{\min }^d}} - p_j^c\bar g_j^s + \xi _s^1\frac{{p_s^d\hat g_s^d}}{{\Gamma _{\min }^d}} - \xi _s^2p_j^c\hat g_j^s,
\end{eqnarray}
where $\xi _s^1 = \frac{{g_s^d - \bar g_s^d}}{{\hat g_s^d}}$, $\xi _s^2 = \frac{{g_j^s - \bar g_j^s}}{{\hat g_j^s}}$. $\xi _s^1$ and $\xi _s^{2}$ are i.i.d. for each $s \in {\cal S}$ and in the range of $[ - 1,1]$.
The values of $\xi _s^1$ and $\xi _s^2$ are independent of each other and belong to the particular class of ${{\cal F}_s}$, which denotes the probability distribution function of $\boldsymbol{\theta} _s^d$.
Then, the intractable chance constraint is transformed into a convex constraint from the following theorem.

\begin{theorem}
Consider a chance constraint as follows
\begin{eqnarray}
{\rm Pr}_{{\xi _l} \sim P} \left\{ {{f_0}(y) + \sum\limits_{l = 1}^d {{\xi _l}{f_l}(y)}  \le 0} \right\} \ge 1 - \beta,\label{2jfgj4r1}
\end{eqnarray}
where ${f_l}(y)$ is the real-valued function of the variable $y$, and the uncertain data ${\xi _l}$ is random with a partially known probability distribution $P$,
so the explicit convex constraint
\begin{align}
{f_0}(y) &+ \sum\limits_{l = 1}^d {\max \left[ {\mu _l^ - {f_l}(y),\mu _l^ + {f_l}(y)} \right]}  \nonumber \\
&+ \sqrt {2\ln (1/\beta)} {\left( {\sum\limits_{l = 1}^d {\sigma _l^2f_l^2(y)} } \right)^{1/2}} \le 0 \label{2jf234e}
\end{align}
is a safe approximation of (\ref{2jfgj4r1}).
\end{theorem}
\proof
Refer to Section IV in \cite{article3}.
\endproof

Therefore, Bernstein approximations of the chance constraint can take the place of constraint (\ref{7c}), i.e.,
\begin{align}
&p_j^c\bar g_j^s - \frac{{p_s^d\bar g_s^d}}{{\Gamma _{\min }^d}} + {\sigma ^2} + \max \left\{ {\mu _{{{\cal F}_s}}^{2 - }p_j^c\hat g_j^s,\mu _{{{\cal F}_s}}^{2 + }p_j^c\hat g_j^s} \right\} \nonumber\\
&+ \max \left\{ {\mu _{{{\cal F}_s}}^{1 - }\frac{{ - p_s^d\hat g_s^d}}{{\Gamma _{\min }^d}},\mu _{{{\cal F}_s}}^{1 + }\frac{{ - p_s^d\hat g_s^d}}{{\Gamma _{\min }^d}}} \right\}\label{2gjgjiije}\\
&+ \sqrt {2\ln \left( {\frac{1}{\beta }} \right)} {\left( {{{\left( {\sigma _{{{\cal F}_s}}^2p_j^c\hat g_j^s} \right)}^2} + {{\left( {\sigma _{{{\cal F}_s}}^1\frac{{ - p_s^d\hat g_s^d}}{{\Gamma _{\min }^d}}} \right)}^2}} \right)^{1/2}} \le 0,\nonumber
\end{align}
where $\sigma _{{{\cal F}_s}}^{1 } \ge 0$, $\sigma _{{{\cal F}_s}}^{2} \ge 0$, $ - 1 \le \mu _{{{\cal F}_s}}^{1 - } \le \mu _{{{\cal F}_s}}^{1 + } \le 1$, $ - 1 \le \mu _{{{\cal F}_s}}^{2 - } \le \mu _{{{\cal F}_s}}^{2 + } \le 1$.
These parameters serve as the safe approximations of the chance constraint and rely on ${{\cal F}_s}$.
Because the parameters including the transmit power, the channel gain and the SINR threshold are positive, we discover
$\mu _{{{\cal F}_s}}^{2 - }p_j^c\hat g_j^s \le \mu _{{{\cal F}_s}}^{2 + }p_j^c\hat g_j^s$ and $\mu _{{{\cal F}_s}}^{1 - }\frac{{ - p_s^d\hat g_s^d}}{{\Gamma _{\min }^d}} \ge \mu _{{{\cal F}_s}}^{1 + }\frac{{ - p_s^d\hat g_s^d}}{{\Gamma _{\min }^d}}$. Moreover, from ${\left( {\sum\limits_{a = 1}^A {{{\left( {{x_a}} \right)}^2}} } \right)^{1/2}} \le \sqrt A \mathop {\max }\limits_{\forall a \in A} \left| {{x_a}} \right|$, we can further get an upper bound of (\ref{2gjgjiije}), i.e.,
\begin{eqnarray}
p_j^c\bar g_j^s - \frac{{p_s^d\bar g_s^d}}{{\Gamma _{\min }^d}} + {\sigma ^2} + \mu _{{{\cal F}_s}}^{2 + }p_j^c\hat g_j^s + \mu _{{{\cal F}_s}}^{1 - }\frac{{ - p_s^d\hat g_s^d}}{{\Gamma _{\min }^d}} \nonumber\\
 + \sqrt {4\ln \left( {\frac{1}{\beta }} \right)} \max \left\{ {\sigma _{{{\cal F}_s}}^2p_j^c\hat g_j^s,\sigma _{{{\cal F}_s}}^1\frac{{p_s^d\hat g_s^d}}{{\Gamma _{\min }^d}}} \right\} \le 0. \label{2gfdfhkiuye}
\end{eqnarray}
Then, (\ref{2gfdfhkiuye}) can be rewritten as
\begin{align}
&\frac{{p_s^d\bar g_s^d}}{{\Gamma _{\min }^d}} - p_j^c\bar g_j^s + \mu _{{{\cal F}_s}}^{1 - }\frac{{p_s^d\hat g_s^d}}{{\Gamma _{\min }^d}} - \mu _{{{\cal F}_s}}^{2 + }p_j^c\hat g_j^s  \nonumber\\
& + \sqrt {4\ln \left( {\frac{1}{\beta }} \right)} \min \left\{ { - \sigma _{{{\cal F}_s}}^2p_j^c\hat g_j^s, - \sigma _{{{\cal F}_s}}^1\frac{{p_s^d\hat g_s^d}}{{\Gamma _{\min }^d}}} \right\} \ge {\sigma ^2}. \label{2vrffdsfse}
\end{align}

From \cite{article4,etsi2018103}, $\mu _{{{\cal F}_s}}^ + $ and $\sigma _{{{\cal F}_s}} $ are relevant to the mean and variance of ${{\cal F}_s}$ as shown in Table I, respectively. Therefore, by applying Bernstein approximations, the chance constraint in (\ref{7c}) is translated into the computable constraint in (\ref{2vrffdsfse}).
\begin{table}[t]
\caption{$\mu _{{{\cal F}_s}}^ + $ and $\sigma _{{{\cal F}_s}} $ for Typical Families of ${{\cal F}_s}$}
\begin{tabular}{lccc}
\hline
${{\cal F}_s}$ is & $\mu _{{{\cal F}_s}}^ - $ & $\mu _{{{\cal F}_s}}^ + $ & $\sigma _{{{\cal F}_s}} $ \\ \hline
\uppercase\expandafter{\romannumeral1}: $\sup \{ {{\cal F}_s}\}  \in [ - 1,1]$ & -1 &1 & 0 \\
\uppercase\expandafter{\romannumeral2}: $\sup \{ {{\cal F}_s}\} $ is uni-modal and $\sup \{ {{\cal F}_s}\}  \in [ - 1,1]$ & $ - \frac{1}{2}$ & $\frac{1}{2}$ & $\frac{1}{{\sqrt {12} }}$ \\
\uppercase\expandafter{\romannumeral3}: $\sup \{ {{\cal F}_s}\} $ is uni-modal and symmetric & 0 & 0 & $\frac{1}{{\sqrt {3} }}$\\\hline
\end{tabular}
\vspace{-1em}
\end{table}

\subsection{Bisection Search Algorithm}

According to above Bernstein approximations for the V2V chance constraint, problem (7) is reformulated as
\begin{align}
C_{j,s} = \max \limits_{\{ p_j^c\} \{ p_s^d\} } &B{\log _2}(1 + \frac{{p_j^cg_j^c}}{{{\sigma ^2} + p_s^dg_s^B}}) \label{13fdfdf}\\
\textrm{s.t.}\quad
&   \textrm{(\ref{8})},\textrm{(\ref{2vrffdsfse})},\textrm{(\ref{7d})}. \nonumber
\end{align}
As discussed above, the V2V chance constraint becomes a tractable constraint. However, problem (\ref{13fdfdf}) is non-convex because of the fractional form of $p_j^c$ and $p_s^d$ in the objective function. Thus, Bisection search algorithm is designed to solve problem (\ref{13fdfdf}). First, either $p_j^c$ or $p_s^d$ should be fixed to search the corresponding optimal solution $p_s^{d,*}$ or $p_j^{c,*}$. Then, the value of the fixed variable is adjusted until both $p_j^c$ and $p_s^d$ are close to the optimal solution of problem (\ref{13fdfdf}).

Based on the above analysis, we choose to fix $p_s^d$ and consider $p_j^c$ as the variable. Thus, the problem is formulated as
\begin{align}
C_{j,s} = \max\limits_{\{ p_j^c\} } \quad&B{\log _2}(1 + \frac{{p_j^cg_j^c}}{{{\sigma ^2} + p_s^dg_s^B}}) \label{djfdjfidj}\\
\textrm{s.t.}\quad\
&\textrm{(\ref{8})},\textrm{(\ref{2vrffdsfse})},0 \le p_j^c.\nonumber
\end{align}
Due to the form of $min\{\}$ in constraint \textrm{(\ref{2vrffdsfse})}, problem (\ref{djfdjfidj}) is first divided into following two cases: (i) Assuming that ${\frac{{\sigma _{{{\cal F}_s}}^{1 }p_s^d\hat g_s^d}}{{\Gamma _{\min }^d}} \leq \sigma _{{{\cal F}_s}}^{2 }p_j^c\hat g_j^s}$, constraint \textrm{(\ref{2vrffdsfse})} can be transformed into a linear constraint so problem (\ref{djfdjfidj}) becomes a concave problem; (ii) Assuming that ${\frac{{\sigma _{{{\cal F}_s}}^{1 }p_s^d\hat g_s^d}}{{\Gamma _{\min }^d}} \geq \sigma _{{{\cal F}_s}}^{2}p_j^c\hat g_j^s}$, constraint \textrm{(\ref{2vrffdsfse})} also becomes a linear constraint and problem (\ref{djfdjfidj}) becomes a concave problem.
Then, the optimal solution $p_j^{c,*}$ of problem (\ref{djfdjfidj}) can be obtained by the interior method.
Finally, we test whether the optimal solutions are inconsistent with the assumptions in both cases, and then discard the contradictory case.

Furthermore, it is noted that if the transmit power $p_j^c$ becomes smaller, the smaller transmit power $p_s^d$ will be given for satisfying the QoS constraint in (7c). It means that $p_s^d$ is a monotone increasing function of $p_j^c$ and it can be expressed as $p_s^d(p_j^c)$. Therefore, Bisection search algorithm can be employed to obtain the approximate optimal solutions $p_j^{c,*}$ and $p_s^{d,*}$ of problem (\ref{13fdfdf}) by constantly adjusting $p_s^d$. According to the conclusion in the following Section V.A, the optimal transmit power of problem (\ref{13fdfdf}) always satisfies either $p_j^{c,*} = p_{\max }^c$ or $p_s^{d,*} = p_{\max }^d$.
Thus, the termination condition of the algorithm can be obtained based on this conclusion.

Consequently, the optimal solution of problem (\ref{13fdfdf}) is resolved by \textbf{Algorithm 1}. Then, the computational complexity of \textbf{Algorithm 1} will be analyzed.
It can be seen that the maximum iteration is eventually decided by line 1 and line 7. The optimal solution $p_s^{d,*}$ can be obtained from the interval $[0,p_{\max }^d]$ and the search region can be cut down by half at each iteration. Therefore, after finishing $N$ iterations, the length of interval is only $p_{\max }^d/{2^n}$. From the above conclusions, the maximum iteration number of \textbf{Algorithm 1} is ${\log _2}(p_{\max }^d/\xi )$, where $\xi$ is the termination threshold.
Because the interior point algorithm is applied to resolve problem (\ref{djfdjfidj}), the complexity of problem (\ref{djfdjfidj}) is ${\cal O}({\rm{log}}\frac{1}{\iota })$, where $\iota$ is the target accuracy \cite{4b7271b331194618b68420852d0a6c66}.
Therefore, the complexity of \textbf{Algorithm 1} is ${\cal O}({\rm{log}}\frac{1}{\iota }{\log _2}(\frac{{p_{\max }^d}}{\xi }))$.
\begin{algorithm}[h]
\caption{Bisection Search for Solving Problem (\ref{13fdfdf})}
\begin{algorithmic}[1]
\REQUIRE $p_{s,min}^{d}=0$, $p_{s,max}^{d}=P_{max}^{d}$ and the termination threshold $0<\xi<1$;
\WHILE {$p_{s}^{d}<P_{max}^{d}-\xi$}
\STATE set $p_{s}^{d}=(p_{s,min}^{d}+p_{s,max}^{d})/2$; Solve problem (\ref{djfdjfidj}) to obtain $p_{j}^{c}$;
\IF{$p_{j}^{c}>P_{max}^{c}+\xi$}
\STATE $p_{s,max}^{d}=p_{s}^{d}$
\ELSIF{$p_{j}^{c}<P_{max}^{c}-\xi$}
\STATE $p_{s,min}^{d}=p_{s}^{d}$
\ELSIF{$P_{max}^{c}-\xi<p_{j}^{c}<P_{max}^{c}+\xi$}
\STATE break
\ENDIF
\ENDWHILE
\ENSURE The optimal solution $p_{j}^{c,*}$ and $p_{s}^{d,*}$.
\end{algorithmic}
\end{algorithm}

\subsection{Spectrum Allocation}

We can acquire the power allocation of all possible spectrum reusing pairs by means of the method in above subsection.
In practical V2X networks, the number of VUE pairs is no larger than CUEs ($S \leq J$) \cite{8638940}. Because each VUE can reuse the spectrum of at most one CUE and the spectrum of each CUE can be reused by at most one VUE.
For the case $J > S$, there is a situation where the spectrums of $J-S$ CUEs can not be reused by VUEs. For considering the spectrum allocation as a bipartite matching problem, a set of virtual VUEs represented as ${\cal S}'$ is constructed as ${\cal S}' = \{ S + 1,S + 2, \cdots ,J\} $.
Since the $s$th VUE $(s \in {\cal S}')$ can not reuse the spectrum of the $j$th CUE $(j \in {\cal J})$, ${p_{\max }^c}$ is allowed as the transmit power of the $j$th CUE and the transmit power of the $s$th VUE $(s \in {\cal S}')$ is regarded as 0.
Therefore, the optimal solution to the power allocation is calculated as ${C_{j,s}} = {B_c}{\log _2}\left( {1 + \frac{{p_c^{\max }g_j^c}}{{{\sigma ^2}}}} \right)$.
After solving the power allocation problem under all possible spectrum pairs, we can find the optimal the spectrum reusing pattern based on the obtained optimal power allocation solutions.
Then, in order to search the optimal spectrum reusing pairs, a combinational optimization problem can be described as
\begin{align}
\max \quad &\sum\limits_{j \in {\cal J}} {\sum\limits_{s \in \{{\cal S \cup {\cal S}'} \} } {{\rho_{j,s}}{C_{j,s}}} } \label{d456ji}\\
\textrm{s.t.}\quad&\sum\limits_{j \in {\cal J}} {{\rho_{j,s}} \le 1} ,\forall s \in \{{\cal S \cup {\cal S}'}\},\nonumber \\
&\sum\limits_{s \in \{{\cal S \cup {\cal S}'} \} } {{\rho_{j,s}} \le 1}, \forall j \in {\cal J},\nonumber \\
&{\rho_{j,s}}  \in \{ 0,1\} , \forall j,s, \nonumber
\end{align}
which is a maximum weight bipartite matching problem and can be efficiently solved by Hungarian method in polynomial time \cite{liang2017spectrum}.
According to the spectrum allocation above, problem (6) can be solved in \textbf{Algorithm 2}.
We have known that the complexity of Hungarian method is ${\cal O}({{\rm{J}}^3})$. Therefore, the complexity of \textbf{Algorithm 2} is ${\cal O}({{\rm{J}}^3}{\rm{ + J R log}}\frac{1}{\iota }{\log _2}(\frac{{p_{\max }^d}}{\xi }))$.
\begin{algorithm}[h]\label{Am1}
\caption{Bernstein Approximation-based Robust Resource Allocation}
\begin{algorithmic}[1]
\REQUIRE The i.i.d. channel state samples of V2I communication and V2V communication;
\FOR {$j=1,\cdots,J$}
\FOR {$s=1,\cdots,S$}
\STATE Obtain $\mu _{{{\cal F}_s}}^ - $, $\mu _{{{\cal F}_s}}^ + $ and $\sigma _{{{\cal F}_s}}$ of ${{\cal F}_s}$ from Table $\uppercase\expandafter{\romannumeral1}$;
\STATE Use a convex function to replace (\ref{7c}) as the safe approximations based on the Bernstein approach;
\STATE Acquire the optimal power solution $(p_{j}^{c,*}, p_{s}^{d,*})$ from \textbf{Algorithm 1};
\ENDFOR
\ENDFOR
\STATE  Apply Hungarian algorithm to compute the optimal spectrum reusing pattern $\{\rho_{j,s}^*\}$ based on $\{C_{j,s}\}$;
\ENSURE The optimal resource allocation $\{\rho_{j,s}^*\}\{p_{j}^{c,*}\}\{p_{s}^{d,*}\}$.
\end{algorithmic}
\end{algorithm}

\section{Self-learning Robust Resource Allocation Approach}

As described on the previous sections, the resource allocation in section IV is achieved by using Bernstein Approximation-based Robust Resource Allocation approach.
Based on the above analysis, though the computation complexity of this approach is not significantly high, its complexity is still high for the real-time high-mobility vehicle network and it is not suitable for the practical environment of V2X.
Therefore, a new approach with lower complexity is proposed to solve problem (6). In the following section, we continue decomposing problem (6) into a power allocation problem and a spectrum allocation problem.

\subsection{Power Allocation}
According to (8) and (9), the power allocation subproblem (7) is reformulated as
\begin{subequations}
\begin{align}
\!\!\!\!\!\!\!\!\!\!\!\!\!C_{j,s} = \max \limits_{\{ p_j^c\} \{ p_s^d\} } &B{\log _2}(1 + \frac{{p_j^cg_j^c}}{{{\sigma ^2} + p_s^dg_s^B}}) \label{dsfedfsedfga}\\
\textrm{s.t.}\quad
&  \Pr \{ \boldsymbol{p}_s^d\boldsymbol{\theta} _s^d \ge {\sigma ^2}\}  \ge 1 - \beta , \label{dsfedfsedfgc} \\
&   \textrm{(8)}, \textrm{(7d)}.\label{dsfedfsedfgb}
\end{align}
\end{subequations}
For transforming the intractable chance constraints into computable ones, we present a learning method which uses a high-probability-region (HPR) to represent the channel uncertainty. Let $\boldsymbol{G}_s^d$ denote the HPR which requires to be learned from the data samples. When $\boldsymbol{G}_s^d$ is chosen to cover a $1 - \beta $ content of $\boldsymbol{\theta} _s^d$, i.e. $\Pr \{ \boldsymbol{\theta} _s^d \in \boldsymbol{G}_s^d\}  \ge 1 - \beta $, the feasible solutions of problem (20) certainly satisfy
\begin{eqnarray}
\Pr \{ \boldsymbol{p}_s^d\boldsymbol{\theta} _s^d \ge {\sigma ^2}\}  \ge \Pr \{ \boldsymbol{\theta} _s^d \in \boldsymbol{G}_s^d\}. \label{10}
\end{eqnarray}
Because of the unknown prior distribution of $\boldsymbol{\theta} _s^d$, the confidence level $1 - \varsigma $ is utilized as the statistic assurance of the robust optimization approach which can satisfy $1 - \beta $ content of $\boldsymbol{\theta} _s^d$. Then, the basic idea is to satisfy (\ref{10}) through constructing $\boldsymbol{G}_s^d$ under the confidence level $1 - \varsigma $, i.e.
\begin{eqnarray}
\Pr \{ \Pr \{ \boldsymbol{\theta} _s^d \in \boldsymbol{G}_s^d\}  \ge 1 - \beta \}  \ge 1 - \varsigma . \label{19}
\end{eqnarray}
Multiple samples of the uncertain CSI $\boldsymbol{\theta} _s^d$ should be collected to learn the uncertainty set so that the distribution of $e_s^d$ and $e_j^s$ can be acquired.
In the sampling process, we consider the block fading channel where the channel distribution is assumed invariant \cite{liu2019two} within $T_B$ blocks, where each block is called a coherence interval for CSI and the large-scale slow fading information remains unchanged within each block.
Therefore, the gNB can collect $N$ i.i.d. samples of the imperfect CSI as the sample set $\boldsymbol{{\rm X}} = \{ {\boldsymbol{\chi} ^{1}},{\boldsymbol{\chi} ^{2}}, \cdots ,{\boldsymbol{\chi} ^{N}}\} $ within each block, where ${\boldsymbol{\chi} ^{n}} = \{ \boldsymbol{\chi} _1^{n},\boldsymbol{\chi} _2^{n}, \cdots ,\boldsymbol{\chi} _S^{n}\} $ and $\boldsymbol{\chi} _{s}^{n}\in {{\mathbb{R}}^{2}}$.
In order to intuitively express the sampling process, the timeline of sampling process is expressed as shown in Fig. 2.

\begin{figure}
\begin{center}
\includegraphics[scale = 0.6]{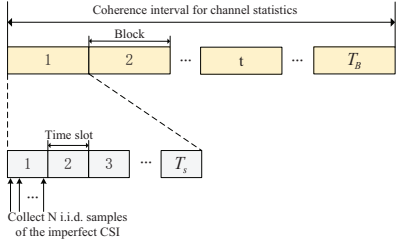}
\caption{Timeline of the sampling process.}\label{dfgdf}
\end{center}
\vspace{-2em}
\end{figure}

Based on the above analysis, the learning method, which includes the set construction and the size calibration, is proposed to determine the size of the HPR.
\subsubsection{Set Construction}

In general, ellipsoid sets are utilized to model the channel uncertainty. However, the robust resource allocation problem containing ellipsoid sets is often transformed into a second-order cone program (SOCP) which is a complicated problem.
Thus, to simplify the problem, the affine sets can be constructed to model the channel uncertainty, such as $\boldsymbol{A}^T\boldsymbol{Y}\ge r$, where $\boldsymbol{A} \in {{\mathbb{R}}^{n}}, \boldsymbol{Y} \in {{\mathbb{R}}^{n}}$ and $r \in {\mathbb{R}}$.
Then, when an initial feasible solution $\boldsymbol{\tilde p}_s^d$ is given, the uncertainty set $\boldsymbol{G}_s^d$ is constructed as
$\boldsymbol{G}_s^d = \{ \boldsymbol{\theta} _s^d|\boldsymbol{\tilde p}_s^d\boldsymbol{\theta} _s^d \ge {r_d}\}$, where ${{r}_{d}}\in \mathbb{R}$ is the size of the HPR $\boldsymbol{G}_s^d$.

\subsubsection{Size Calibration}
$\boldsymbol{G}_s^d$ is calibrated by the sample set $\boldsymbol{{\rm X}}$ to satisfy (\ref{19}).
The quantile evaluation approach, which is the critical idea of size calibration, is utilized to measure the size of the affine uncertainty set \cite{9382930}.
Thus, the $1-\beta$ quantile $t_{1-\beta}$ with $1-\varsigma$ confidence of data samples in $\boldsymbol{{\rm X}}$ can be estimated to calibrate the uncertainty set $\boldsymbol{G}_s^d$.
First, let
${f\boldsymbol{_\chi} }({\boldsymbol{\chi} ^{n}}) = \mathop {\min }\limits_{s = 1, \cdots ,S} \{ \boldsymbol{\tilde p}_s^d\boldsymbol{\chi} _s^{n}\}$
be the mapping from the stochastic space ${{\mathbb{R}}^{2S}}$ into $\mathbb{R}$.
$\boldsymbol{G}_s^d$ demonstrates that the value of $Pr \{ \boldsymbol{\theta} _s^d \in \boldsymbol{G}_s^d\}$ should be larger than $1-\beta$ based on $1-\varsigma$ confidence.
Second, $t_{1-\beta}$ of the potential distribution of ${f}({\boldsymbol{\chi} ^{(n)}})$ is defined under the sample set of $\boldsymbol{{\rm X}}$ based on $Pr\{{f}({\boldsymbol{\chi} ^{(n)}})\leq t_{1-\beta}\}=1-\beta$.
Thus, we can obtain ${f\boldsymbol{_\chi} }({\boldsymbol{\chi} ^{(k_{\boldsymbol{{\chi}}}^*)}})$ which is the $1-\varsigma$ confidence upper boundary in regard to $t_{1-\beta}$ from the following theorem.
\begin{theorem}
${f\boldsymbol{_\chi} }({\boldsymbol{\chi} ^{(1)}})  \leq {f\boldsymbol{_\chi} }({\boldsymbol{\chi} ^{(2)}}) \leq \cdots \leq  {f\boldsymbol{_\chi} }({\boldsymbol{\chi} ^{(N)}})$
is defined as the order statistics based on the samples of $\boldsymbol{{\rm X}}$.
${f\boldsymbol{_\chi} }({\boldsymbol{\chi} ^{(k_{\boldsymbol{{\chi}}}^*)}})$ can be obtained from $$Pr\{r_d \geq t_{1-\beta}\}=1-\varsigma,$$ where
$k_{\boldsymbol{{\chi}}}^*$ is computed as
$$\min \Bigg\{ {{k\boldsymbol{_\chi}}:\sum\limits_{t = 0}^{{k\boldsymbol{_\chi}} - 1} { \Bigg ( \begin{array}{l}
N\\
\,t
\end{array} \Bigg){{(1 - \beta )}^t}{\beta ^{N - t}} \ge 1 - \varsigma } } \Bigg\}.$$
\end{theorem}
\proof
Refer to Theoren 1 in \cite{9382930}.
\endproof

Third, the size of ${r_d}$ can be given as
\begin{eqnarray}
{r_d} = {f\boldsymbol{_\chi} }({\boldsymbol{\chi} ^{_{(k\boldsymbol{_\chi} ^*)}}}). \nonumber
\end{eqnarray}
Based on the uncertainty set $\boldsymbol{G}_s^d$, problem (20) is transformed into
\begin{subequations}
\begin{align}
\!\!\!\!\!\!\!\!\!\!\!\!\!C_{j,s} = \max \limits_{\{ p_j^c\} \{ p_s^d\} } &B{\log _2}(1 + \frac{{p_j^cg_j^c}}{{{\sigma ^2} + p_s^dg_s^B}}) \label{dsfwerredfsedfga}\\
\textrm{s.t.}\quad
&  \boldsymbol{p}_s^d\boldsymbol{\theta} _s^d \geq \sigma^2, \boldsymbol{\theta} _s^d \in \boldsymbol{G}_s^d, \label{dsfedfsdgsafge} \\
&   \textrm{(8)}, \textrm{(7d)}.\label{dsfvfgrgredfsedfgb}
\end{align}
\end{subequations}
According to constraint (\ref{dsfedfsdgsafge}) for V2V communication, the tractable QoS constraints are obtained by the following optimization problem
\begin{align}
\mathop {\min }\limits_{\boldsymbol{\theta} _s^d} \quad &\boldsymbol{p}_s^d\boldsymbol{\theta} _s^d\label{15}\\
\textrm{s.t.}\quad
&\boldsymbol{\tilde p}_s^d\boldsymbol{\theta} _s^d \ge {r_d}. \nonumber
\end{align}
The dual problem of problem (\ref{15}) is
\begin{align}
\quad\quad\quad\quad\mathop {\max} \limits_{z_{\rm{s}}^d} \quad &z_{\rm{s}}^d{r_d}\\
\textrm{s.t.}\quad
&z_{\rm{s}}^d\boldsymbol{\tilde p}_s^d \le \boldsymbol{p}_s^d,z_{\rm{s}}^d \ge 0. \nonumber
\end{align}
From above analysis, the chance constraint in (\ref{dsfedfsedfgc}) is transformed into following constraints
 \begin{subequations}  
\begin{numcases}{}
z_{\rm{s}}^d{r_d} \geq \sigma^2,\\
z_{\rm{s}}^d\boldsymbol{\tilde p}_s^d \le \boldsymbol{p}_s^d,z_{\rm{s}}^d \ge 0. \label{fsbsedte}
\end{numcases}
\end{subequations}
Thus, problem (20) becomes
\begin{subequations}
\begin{align}
\!\!\!\!\!\!\!\!\!\!\!\!\!C_{j,s} = \max \limits_{\{ p_s^d\} {\rm{\} \{ }}p_j^c {\rm{\{ }}z_{\rm{s}}^d{\rm{\} }}} &B{\log _2}(1 + \frac{{p_j^cg_j^c}}{{{\sigma ^2} + p_s^dg_s^B}})\label{17a}\\
\textrm{s.t.}\quad\
& z_{\rm{s}}^d\boldsymbol{\tilde p}_s^d \le \boldsymbol{p}_s^d,\label{17d}\\
& z_{\rm{s}}^d{r_d} \ge {\sigma ^2},z_{\rm{s}}^d \ge 0, \textrm{(8)},\textrm{(7d)}. \label{17c}
\end{align}
\end{subequations}
Because the variables $p_j^c$ and $ p_s^d$ coexist in the numerator and denominator of the objective function in (\ref{17a}), problem (27) is still non-convex. For overcoming this challenge, the following analytical mapping method is proposed to resolve problem (27).

To clearly reflect the relationships between different constraints, constraint (\ref{17d}) is equivalent to
\begin{subequations}  
\begin{numcases}{}
z_{\rm{s}}^d\tilde p_s^d \le p_s^d, \label{asca}\\
z_{\rm{s}}^d\tilde p_j^c \ge p_j^c. \label{ascb}
\end{numcases}
\end{subequations}
Then, together with constraint (\ref{7d}), the following two simpler constraints are obtained as
\begin{subequations}  
\begin{numcases}{}
0 \leq p_j^c \leq \min \{z_{\rm{s}}^d\tilde p_j^c,p_{\max }^c\}, \label{fada} \\
z_{\rm{s}}^d\tilde p_s^d \le p_s^d \le p_{\max }^d. \label{fadb}
\end{numcases}
\end{subequations}
By comparing the values of $z_{\rm{s}}^d\tilde p_j^c$ and $p_{\max }^c$ in constraint (\ref{fada}), problem (27) is divided into two cases.

\textit{1) Case 1}:
When $z_{\rm{s}}^d\tilde p_j^c \le p_{\max }^c$ and problem (27) is feasible, then $p_s^d$ and $p_j^c$ satisfy
\begin{subequations}  
\begin{numcases}{}
\frac{{p_j^cg_j^c}}{{\Gamma _{\min }^c}}{\rm{ - }}p_s^dg_s^B \ge {\sigma ^2}, \label{asfaa}\\
z_{\rm{s}}^d\tilde p_s^d \le p_s^d \le p_{\max }^d, \label{asfab}\\
0 \leq p_j^c \le z_{\rm{s}}^d\tilde p_j^c \le p_{\max }^c. \label{asfac}
\end{numcases}
\end{subequations}
We can obtain the feasible region as shown in the shadow of Fig. 3, where $E$ is the intersection point of $p_s^d = z_{\rm{s}}^d\tilde p_s^d$ and $p_j^c = z_{\rm{s}}^d\tilde p_j^c$.
The CUE QoS constraint of (\ref{8}) is plotted as $K_c$.
Based on the objective function of problem (27), the following conclusions can be obviously obtained. When $p_s^d$ is fixed, the capacity ${C_{j,s}}$ monotonely increases with $p_j^c$. In contrast, when $p_j^c$ is fixed, the capacity ${C_{j,s}}$ monotonely decreases with $p_s^d$. It is not difficult to find that the optimal solution of problem (27) must be located in the coordinate of $E$ which can be computed as $(z_{\rm{s}}^d\tilde p_j^c,z_{\rm{s}}^d\tilde p_s^d)$.
Moreover, when the feasible region exists, the optimal solution of problem (27) is uncorrelated with the constraint (\ref{asfaa}) because it is an inactive constraint.
Then, according to the coordinate of $E$, the optimal solution of problem (27) only depends on $z_{\rm{s}}^d$ and is not affected by the variables $p_j^c$ and $p_s^d$. Thus, constraints (\ref{asfab}) and (\ref{asfac}) can be transformed into $z_{\rm{s}}^d \le \frac{{p_{\max }^d}}{{\tilde p_s^d}}$ and $z_{\rm{s}}^d \le \frac{{p_{\max }^c}}{{\tilde p_j^c}}$, respectively. Then, by substituting the coordinate of $E$ into problem (27), we obtain
\begin{figure}
\begin{center}
\includegraphics[scale = 0.3]{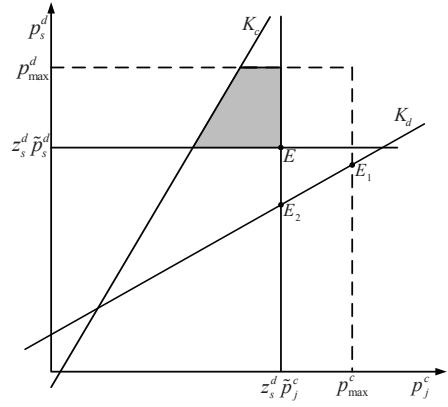}
\label{fig2}\caption{The feasible region of Case 1.}
\vspace{-2em}
\end{center}
\end{figure}
\begin{subequations}
\begin{align}
C_{j,s} = \max \limits_{{\rm{\{ }}z_{\rm{s}}^d{\rm{\} }}} \quad&B{\log _2}(1 + \frac{{z_{\rm{s}}^d\tilde p_j^cg_j^c}}{{{\sigma ^2} + z_{\rm{s}}^d\tilde p_s^dg_s^B}})  \label{sfhsfhsa}\\
\textrm{s.t.}\quad\;
&   \frac{{z_{\rm{s}}^d\tilde p_j^cg_j^c}}{{\Gamma _{\min }^c}}{\rm{ - }}z_{\rm{s}}^d\tilde p_s^dg_s^B \ge {\sigma ^2},\\
&  z_{\rm{s}}^d{r_d} \ge {\sigma ^2}, z_{\rm{s}}^d \ge 0,\\
&  z_{\rm{s}}^d \le \frac{{p_{\max }^d}}{{\tilde p_s^d}}, z_{\rm{s}}^d \le \frac{{p_{\max }^c}}{{\tilde p_j^c}}.
\end{align}
\end{subequations}
By combining the constraints of problem (31), we discover that the feasible region exists when $z_{\rm{s}}^d$ satisfies
\begin{eqnarray}
\max \{ \Upsilon _j^c,{\Delta _d}\}  \le z_{\rm{s}}^d \le \min \{ \Lambda _s^d,\Lambda _j^c\},\label{21}
\end{eqnarray}
where $\Upsilon _j^c = \frac{{{\sigma ^2}\Gamma _{\min }^c}}{{\tilde p_j^cg_j^c - \Gamma _{\min }^c\tilde p_s^dg_s^B}}$, ${\Delta _d} = \frac{{{\sigma ^2}}}{{{r_d}}}$, $\Lambda _s^d = \frac{{p_{\max }^d}}{{\tilde p_s^d}}$ and $\Lambda _j^c = \frac{{p_{\max }^c}}{{\tilde p_j^c}}$. Note that the capacity ${C_{j,s}}$ monotonely increases with $z_{\rm{s}}^d$, so the optimal solution of problem (31) should be $z_{\rm{s}}^{d,*} = \min \{ \Lambda _s^d,\Lambda _j^c\} $. Then, by substituting the value of $z_{\rm{s}}^{d,*}$ into the coordinate of $E$ to obtain $p_j^{c,*}$ and $p_s^{d,*}$, the capacity $C_{j,s}^{*}$ can be calculated. If $z_{\rm{s}}^d$ dissatisfies (\ref{21}), problem (31) is not feasible and the capacity is set as $C_{j,s}^{*} = 0$.

\textit{2) Case 2}:
When $p_{\max }^c \le z_{\rm{s}}^d\tilde p_j^c$ and problem (27) is feasible, then $p_s^d$ and $p_j^c$ satisfy
\begin{subequations}  
\begin{numcases}{}
    \frac{{p_j^cg_j^c}}{{\Gamma _{\min }^c}}{\rm{ - }}p_s^dg_s^B \ge {\sigma ^2}, \label{sfsfsfdfsa}\\
    z_{\rm{s}}^d\tilde p_s^d \le p_s^d \le p_{\max }^d, \label{sfsfsfdfsb}\\
    0 \leq p_j^c \le p_{\max }^c \le z_{\rm{s}}^d\tilde p_j^c. \label{sfsfsfdfsc}
\end{numcases}
\end{subequations}
We can obtain the feasible region as shown in the shadow of Fig. 4, where $F$ is the intersection point of $p_s^d = z_{\rm{s}}^d\tilde p_s^d$ and $p_j^c = p_{\max }^c$. Similar to the conclusion of Case 1, the optimal solution of problem (27) resides at the coordinate of $F$ which is expressed as $(p_{\max }^c,z_{\rm{s}}^d\tilde p_s^d)$. Similarly, if the feasible region exists, the optimal solution of problem (27) is uncorrelated with constraint (\ref{sfsfsfdfsa}).  Meanwhile, constraints (\ref{sfsfsfdfsb}) and (\ref{sfsfsfdfsc}) are combined into $\frac{{p_{\max }^c}}{{\tilde p_j^c}} \le z_{\rm{s}}^d \le \frac{{p_{\max }^d}}{{\tilde p_s^d}}$. Thus, by substituting the coordinate of $F$ into problem (27), it becomes
\begin{subequations}
\begin{align}
C_{j,s} = \max \limits_{{\rm{\{ }}z_{\rm{s}}^d{\rm{\} }}} \quad&B{\log _2}(1 + \frac{{p_{\max }^cg_j^c}}{{{\sigma ^2} + z_{\rm{s}}^d\tilde p_s^dg_s^B}})\\
\textrm{s.t.}\quad\;
&   \frac{{p_{\max }^cg_j^c}}{{\Gamma _{\min }^c}}{\rm{ - }}z_{\rm{s}}^d\tilde p_s^dg_s^B \ge {\sigma ^2},\\
&  z_{\rm{s}}^d{r_d} \ge {\sigma ^2}, z_{\rm{s}}^d \ge 0\\
&  \frac{{p_{\max }^c}}{{\tilde p_j^c}} \le z_{\rm{s}}^d \le \frac{{p_{\max }^d}}{{\tilde p_s^d}}.
\end{align}
\end{subequations}
\begin{figure}
\begin{center}
\includegraphics[scale = 0.3]{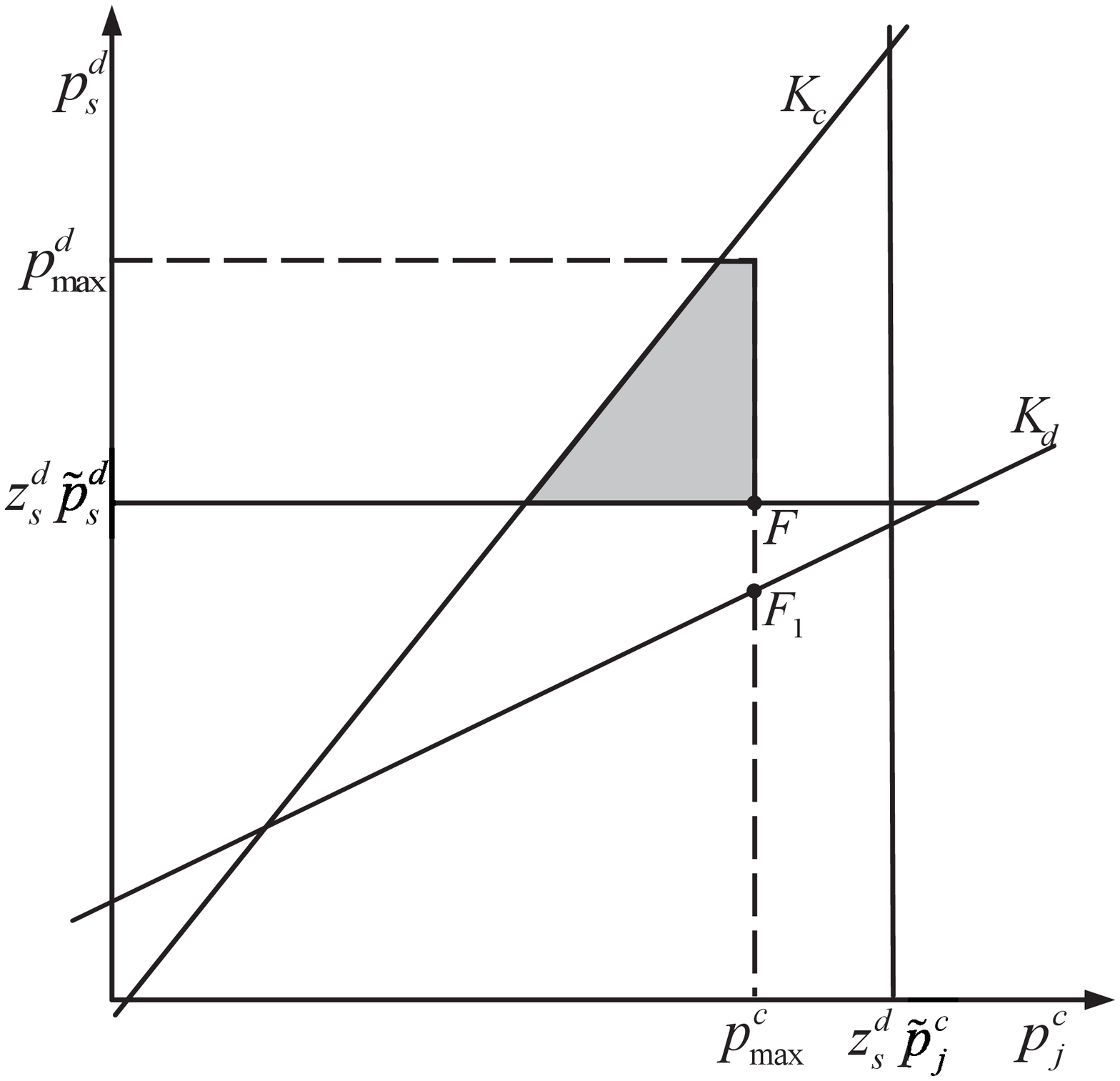}
\label{fig3}\caption{The feasible region of Case 2.}
\end{center}
\vspace{-2em}
\end{figure}
Then, by combining the constraints of problem (34), it will become feasible when $z_{\rm{s}}^d$ satisfies
\begin{align}
\max \{ {\Delta _d},\Lambda _j^c\}  \le z_{\rm{s}}^d \le \min \{ \Omega _j^c,\Lambda _s^d\},\label{24}
\end{align}
where $\Omega _j^c = \frac{{p_{\max }^cg_j^c - {\sigma ^2}\Gamma _{\min }^c}}{{\Gamma _{\min }^c\tilde p_s^dg_s^B}}$. Further analysis about the objective function of problem (34) shows that the capacity ${C_{j,s}}$ monotonely decreases with $z_{\rm{s}}^d$. Therefore, the optimal solution of problem (34) is $z_{\rm{s}}^{d,*} = \max \{ {\Delta _d},\Lambda _j^c\} $. Then, we can substitute $z_{\rm{s}}^{d,*}$ into the coordinate of $F$ to obtain $p_j^{c,*}$ and $p_s^{d,*}$ so that the capacity $C_{j,s}^{*}$ can be figured out. If $z_{\rm{s}}^d$ dissatisfies (\ref{24}), problem (34) is not feasible and the capacity is set as $C_{j,s}^{*} = 0$.

To sum up, the power allocation solution can be acquired as
\begin{eqnarray}
(p_j^{c,*},p_s^{d,*}) = \arg \mathop {\max }\limits_{(p_j^c,p_s^d) \in {\cal C}} \,\left\{ B_c{\log _2}(1 + \frac{{p_j^cg_j^c}}{{{\sigma ^2} + p_s^dg_s^B}})\right\},\label{25}
\end{eqnarray}
where
\begin{small}
\begin{equation}
C=\left\{
\begin{aligned}
\nonumber
\{ (p_{\max }^c,\frac{{p_{\max }^c\tilde p_s^d}}{{\tilde p_j^c}})\} , &\textrm{if}\max \{ \Upsilon _j^c,{\Delta _d}\}  \le \Lambda _j^c \le \min \{ \Omega _j^c,\Lambda _s^d\} \\
\{ (\frac{{p_{\max }^d\tilde p_j^c}}{{\tilde p_s^d}},p_{\max }^d)\} , &\textrm{if}\max \{ \Omega _j^c,{\Delta _d}\}  \le \Lambda _s^d \le \Lambda _j^c\\
\{ (p_{\max }^c,\frac{{{\sigma ^2}\tilde p_s^d}}{{{r_d}}})\} ,\quad &\textrm{if}\Lambda _j^c \le {\Delta _d} \le \min \{ \Omega _j^c,\Lambda _s^d\} \\
0, \quad\quad\quad\quad\quad\quad\quad&\textrm{otherwise}.
\end{aligned}
\right.
\end{equation}
\end{small}

It is apparent to find that (\ref{25}) only requires several multiplication procedures and has not loop structures. Therefore, the complexity of the power allocation is ${\cal O}(1)$, which is obviously lower than \textbf{Algorithm 1} of Bernstein Approximation-based Robust Resource Allocation approach.

\subsection{Spectrum Allocation}
According to above subsection, we can obtain the optimal solutions $(p_j^{c,*},p_s^{d,*})$ under each possible spectrum reusing pair. For searching the optimal spectrum reusing pattern, Hungarian method is applied to resolve problem (\ref{d456ji}).
Then, Self-learning Robust Resource Allocation approach has been summarized in \textbf{Algorithm 3}.
Because the complexity of the spectrum allocation is ${\cal O}({{\rm{J}}^3})$, the general complexity of \textbf{Algorithm 3} is ${\cal O}({{\rm{J}}^3} + {\rm{JR}})$ \cite{wu2021robust}.

\begin{algorithm}[h]
\caption{Self-learning Robust Resource Allocation}
\label{Am111}
\begin{algorithmic}[1]
\REQUIRE An initial feasible solution $\boldsymbol{\tilde p}_s^d$ and the dada set $\boldsymbol{{\rm X}} = \{ {\boldsymbol{\chi} ^{1}},{\boldsymbol{\chi} ^{2}}, \cdots ,{\boldsymbol{\chi} ^{N}}\} $;
\FOR {$j=1,\cdots, J$}
\FOR {$s=1,\cdots, S$}
\STATE  Compute the values of $\Upsilon _j^c$, $\Omega _j^c$, ${\Delta _d}$, $\Lambda _s^d$ and $\Lambda _j^c$;
\IF {these values cannot satisfy (\ref{21}) or (\ref{24})}
\STATE Set $C_{j,s}=0$;
\STATE \textbf{continue}
\ELSE
\STATE Compute the optimal solution $(p_{j}^{c,*}, p_{s}^{d,*})$ from (\ref{25});
\STATE Obtain the capacity $C_{j,s}$ using $(p_{j}^{c,*}, p_{s}^{d,*})$;
\ENDIF
\ENDFOR
\ENDFOR
\STATE  Apply Hungarian algorithm to compute the optimal spectrum reusing pattern $\{\rho_{j,s}^{*}\}$ based on $\{C_{j,s}\}$;
\ENSURE The optimal resource allocation $\{\rho_{j,s}^{*}\}\{p_{j}^{c,*}\}\{p_{s}^{d,*}\}$.
\end{algorithmic}
\end{algorithm}

\textcolor{blue}{
\section{The Analyses of Suboptimality}}

In this section, we characterize the degrees of suboptimality of \textbf{Algorithm 2} and \textbf{3} by means of the analyses.

\subsection{Suboptimality of Algorithm 2}

We first characterize the degrees of suboptimality of \textbf{Algorithm 2} by means of the following analysis.
When we do not consider the uncertain CSI of the V2V links, the power allocation problem can be formulated as
\begin{subequations}
\begin{align}
C_{j,s} = \max \limits_{\{ p_j^c\} \{ p_s^d\} } &B_c{\log _2}(1 + \frac{{p_j^cg_j^c}}{{{\sigma ^2} + p_s^dg_s^B}})\label{gdfgvbj}\\
\textrm{s.t.}\quad
&   \mathbf{p}_s^d\boldsymbol{\theta} _s^d \ge {\sigma ^2},\label{fhsfuih}\\
&   \textrm{(8)}, \textrm{(7d)}. \label{2gfdfse}
\end{align}
\end{subequations}
We discover that problem (37) is convex and can be obtained by means of the toolbox CVX.

In Section IV, the uncertain CSI of the V2V links is considered and constraint (\ref{2vrffdsfse}) is replaced by Bernstein approximations of the chance constraint, i.e.,
\begin{align}
\mathbf{p}_s^d \boldsymbol{\bar{\theta}} _s^d + \Delta (p_s^d) \ge {\sigma ^2}, \label{fvgvdgsfse}
\end{align}
where $\Delta (p_s^d) = \mu _{{{\cal F}_s}}^{1 - }\frac{{p_s^d\hat g_s^d}}{{\Gamma _{\min }^d}} - \mu _{{{\cal F}_s}}^{2 + }p_j^c\hat g_j^s + \sqrt {4\ln \left( {\frac{1}{\beta }} \right)} \min \left\{ { - \sigma _{{{\cal F}_s}}^1p_j^c\hat g_j^s, - \sigma _{{{\cal F}_s}}^2\frac{{p_s^d\hat g_s^d}}{{\Gamma _{\min }^d}}} \right\}$ is the protection function for problem (7), whose value (called the protection value) is negative and depends on the uncertain parameters.

The optimal capacity of all CUEs for problem (37) and (7) is represented as ${v^*}$ and $v_1^*$, respectively.
Let ${u^*}(\Delta ) = \sup \left\{ {B_c}{{\log }_2}\left( {1 + \frac{{p_j^cg_j^c}}{{{\sigma ^2} + p_s^dg_s^B}}} \right)| \textrm{(\ref{8})}, \textrm{(\ref{fvgvdgsfse})}, \textrm{(\ref{7d})} \right\}.$
When $\Delta$ is small, ${u^*}(\Delta )$ is differentiable with respect to $\Delta$. Using Taylor series, we write ${u^*}(\Delta ) = {u^*}(0) + \Delta \frac{{\partial {u^*}(0)}}{{\partial \Delta }} + o$, where ${u^*}(0)$ is the optimal value for (\ref{gdfgvbj}). Since (\ref{gdfgvbj}) is convex, ${u^*}(\Delta)$ is obtained from the Lagrange dual function of (\ref{gdfgvbj}), i.e., from
\begin{align}
L\left( {\mathbf{p},\boldsymbol{\lambda} } \right) &={B_c}{\log _2}\left( {1 + \frac{{p_j^cg_j^c}}{{{\sigma ^2} + p_s^dg_s^B}}} \right)\nonumber\\
&+ {\lambda _1}\left( {\frac{{p_j^cg_j^c}}{{\Gamma _{\min }^c}} - p_s^dg_s^B - {\sigma ^2}} \right)+ {\lambda _2}\left( {p_s^d\bar \theta _s^d + \Delta  \ge {\sigma ^2}} \right) \nonumber\\
&+ {\lambda _3}\left( {p_{\max }^c - p_j^c} \right) + {\lambda _4}\left( {p_{\max }^d - p_s^d} \right), \label{12fs223e}
\end{align}
and from the sensitivity analysis in \cite{cacuci2005sensitivity}, we have $\frac{{\partial {u^*}(0)}}{{\partial \Delta }} \approx {\lambda _2}$, so $v_1^* - {v^*} \approx {\lambda _2}\Delta $. Since $\lambda _2$ is a nonnegative Lagrange multiplier, the optimal capacity is reduced, as compared with the case in which complete CSI is available.
Hence, the reduction of the optimal capacity is defined by
\begin{align}
{d_1} = {v^*} - v_1^*=-{\lambda _2}\Delta \geq 0, \label{12fdsfe}
\end{align}
which also describes the degrees of suboptimality of \textbf{Algorithm 2}.

\subsection{Suboptimality of Algorithm 3}

Then, we characterize the degrees of suboptimality of \textbf{Algorithm 3} by means of the following analysis.
In order to compare the optimal capacity of problem (37) and problem (27), the feasible regions of both problem (37) and problem (27) are plotted in Fig. 3 and Fig. 4 because they have partially common constraints and feasible regions.

When case 1 is satisfied, we can obtain the feasible region as shown in the shadow of Fig. 3.
The VUE QoS constraint of (\ref{fhsfuih}) are plotted as $K_d$.
$E_1$ is the intersection point of $K_d$ and $p_j^c =p_{max}^c$.
$E_2$ is the intersection point of $K_d$ and $p_j^c = z_{\rm{s}}^d\tilde p_j^c$.
From Section V, we find that the optimal solution of problem (27) must be located in the coordinate of $E$. Likewise, the optimal solution of problem (37) must be located in the coordinate of $E_1$.
We also discover that the capacity of $E_2$ is larger than that of $E$.
By substituting the boundary line function of $K_d$ to the capacity of CUE, we obtain
\begin{align}
{C_{j,s}}\left( {p_j^c} \right) = {\log _2}\left( {1 + \frac{{p_j^cg_j^cg_s^d}}{{g_s^d{\sigma ^2} + \left( {{\sigma ^2} + p_j^cg_j^s} \right)\Gamma _{\min }^dg_s^B}}} \right). \label{fffsfs}
\end{align}
It is not difficult to find that the derivative of ${C_{j,s}}\left( {p_j^c} \right)$ is always greater than zero for $p_j^c\geq 0$. Hence, the function in (\ref{fffsfs}) is an increasing function.
It can be concluded that the capacity of $E_1$ is larger than that of $E_2$ and the capacity of $E_1$ is larger than that of $E$.

When case 2 is satisfied, we can obtain the feasible region as shown in the shadow of Fig. 4.
$F_1$ is the intersection point of $K_d$ and $p_j^c =p_{max}^c$. Similar to the conclusion of Case 1, the optimal solution of problem (27) resides at the coordinate of $F$.
Likewise, the optimal solution of problem (37) must be located in the coordinate of $F_1$.
It is not difficult to find that the optimal capacity for case 2 of problem (27) is smaller than that of problem (37).

Then, let $f({x_0},{y_0})$ is the optimal capacity of problem (37), where $x_0$ and $y_0$ are the optimal power $p_j^{c,*}$ and $p_s^{d,*}$ of problem (37), respectively. Using Taylor series, we write
\begin{align}
f({x_0},{y_0}) = &f({x_1},{y_1}) + ({x_0} - {x_1})f_x^{'}({x_1},{y_1}) \nonumber\\
&+ ({y_0} - {y_1})f_y^{'}({x_1},{y_1}), \label{fffgvb}
\end{align}
where $f({x_1},{y_1})$ is the optimal capacity of problem (27), $x_1$ and $y_1$ are the optimal power $p_j^{c,*}$ and $p_s^{d,*}$ of problem (27), respectively.
Thus, the reduction of the sum capacity of all CUEs is defined by
\begin{align}
{d_2} &= f({x_0},{y_0}) - f({x_1},{y_1}) \nonumber\\
&=({x_0} - {x_1})f_x^{'}({x_1},{y_1}) + ({y_0} - {y_1})f_y^{'}({x_1},{y_1}) \geq 0, \label{fdffgd}
\end{align}
which also describes the degrees of suboptimality of \textbf{Algorithm 3}.
The degrees of suboptimality of \textbf{Algorithm 2} and \textbf{Algorithm 3} are characterized by means of the simulations in Section VII.

\section{Simulation Results}

\begin{table}[t]
\caption{SIMULATION PARAMETERS}
\begin{tabular}{|c|c|}\hline
\textbf{Parameter}&\textbf{Value}\\\hline
        Resource blocks bandwidth, $B_c$       &10 MHz\\\hline
        Carrier frequency, $f_v$             &2 GHz\\\hline
        Noise power, $\sigma^2$              &-174 dBm\\\hline
        Channel feedback delay, $T$        &0.5 ms\\\hline
		Outage probability, $\beta $       &0.05 \\ \hline
        Confidence level, $1-\varsigma$      &0.95\\ \hline
        vehicle speed, $\nu$                 &$80\ {\rm{km/h}}$\\ \hline
        SINR threshold of V2I, $\Gamma _{\min }^c$ &2\\ \hline
        SINR threshold of V2V, $\Gamma _{\min }^d$ &1\\ \hline
        Maximum power of V2I, $p_{\max }^c$  &$30\ {\rm{dBm}}$\\ \hline
        Maximum power of V2V, $p_{\max }^d$  &$30\ {\rm{dBm}}$\\ \hline
        Pathloss model                      &$128.1+37.6log_{10}d$\\ \hline
        Shadowing standard deviation         &8 dB(CUE), 4 dB(VUE)\\ \hline
        Small-scale fast fading              &Rayleigh fading\\ \hline
		Sample number of the sample set, $X$&3000\\\hline
        Sample number of the test set, $X$&6000\\\hline
        Bisection search accuracy, $\xi$ &$10^{-4}$\\ \hline
        Interior point method accuracy, $\iota$ &$10^{-3}$\\ \hline
\end{tabular}
\vspace{-1em}
\end{table}

This section evaluates the performance of the above optimization methods. We consider a single cellular vehicle network as shown in Fig. 1. The gNB models, whose coverage areas are approximately 200 meters, are extensively applied in the simulation of V2X networks \cite{wei2018wireless}.
In addition, it is widely believed that the roads are far away from the gNB. For guaranteeing the gNB to cover all vehicles, we suppose that the distance is at least 100m between the gNB and the roads.
Thus, we assume that the distance between CUEs and the gNB is uniformly distributed in a range of [100 200]. There are 4 CUEs and 4 VUEs in the roads. To ensure the vehicles' security, we consider that the average secure distance between VUEs should be $2.5\sec \times \nu (m/\sec )$.
We apply the complex Gaussian distribution to obtain the evaluated error of $e_s^d$ and $e_j^s$. Moreover, Table II is summarized to enumerate the other significant parameters and the channel information.
In the simulation, the first approach is Bernstein Approximation-based Robust Resource Allocation (BRRA) in Section IV. Then, Self-learning Robust Resource Allocation approach in Section V is simulated by choosing the different initial feasible solutions which are obtained from the worst and the average uncertain CSI, respectively.
We represent them as Self-learning Worst approach (SLWA) and Self-learning Average approach (SLAA). Besides, we simulate a non-robust resource allocation (NRRA), which only considers the large-scale slow fading channel.
In a word, we will compare the four approaches above to evaluate their performances.

\begin{figure}[t]
\centering
\subfigbottomskip=-2pt
\subfigure[]
{
\begin{minipage}[c]{1.0\linewidth}
\centering
\includegraphics[scale = 0.4]{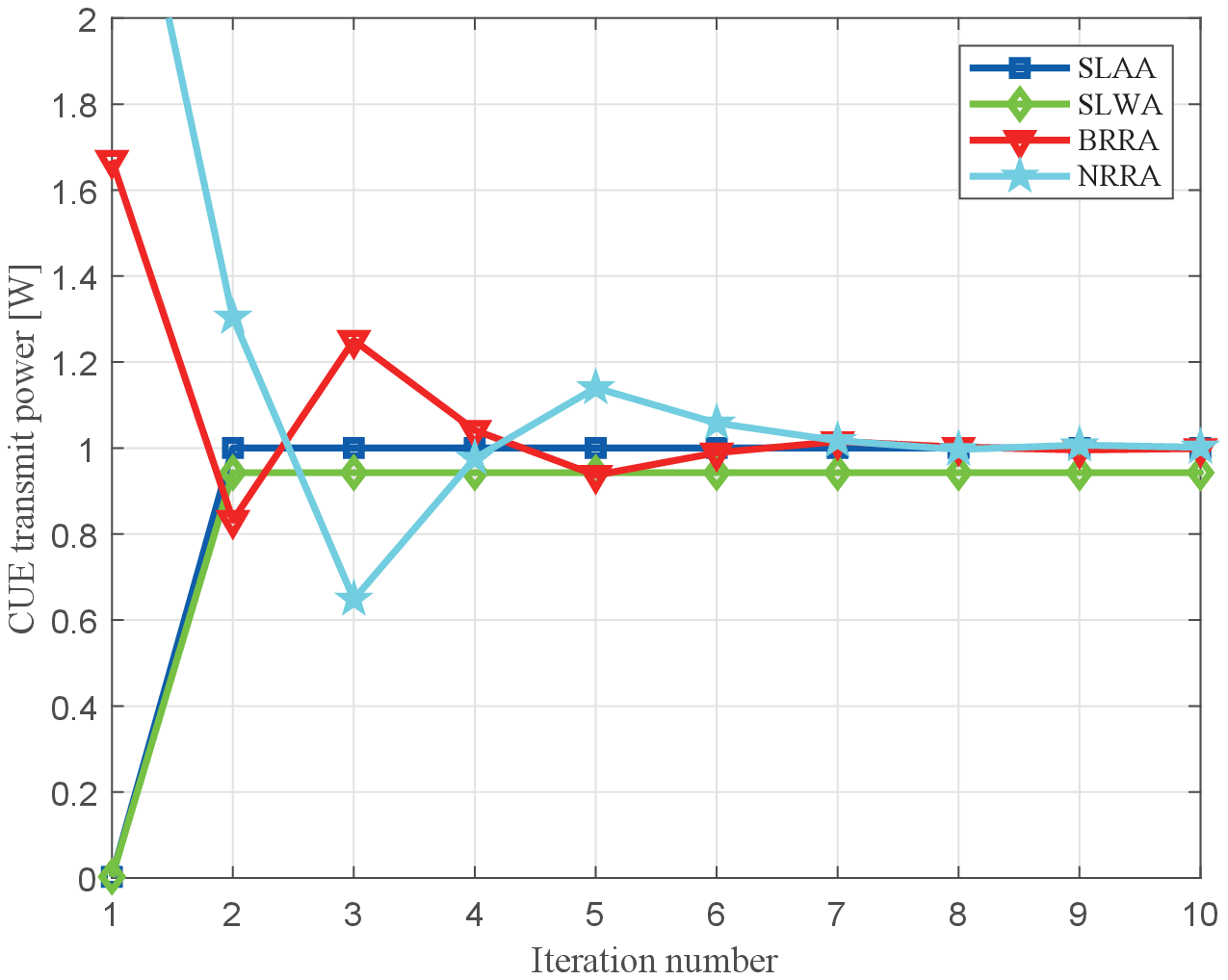}
\end{minipage}%
}\\
\subfigure[]
{
\begin{minipage}[c]{1.0\linewidth}
\centering
\includegraphics[scale = 0.4]{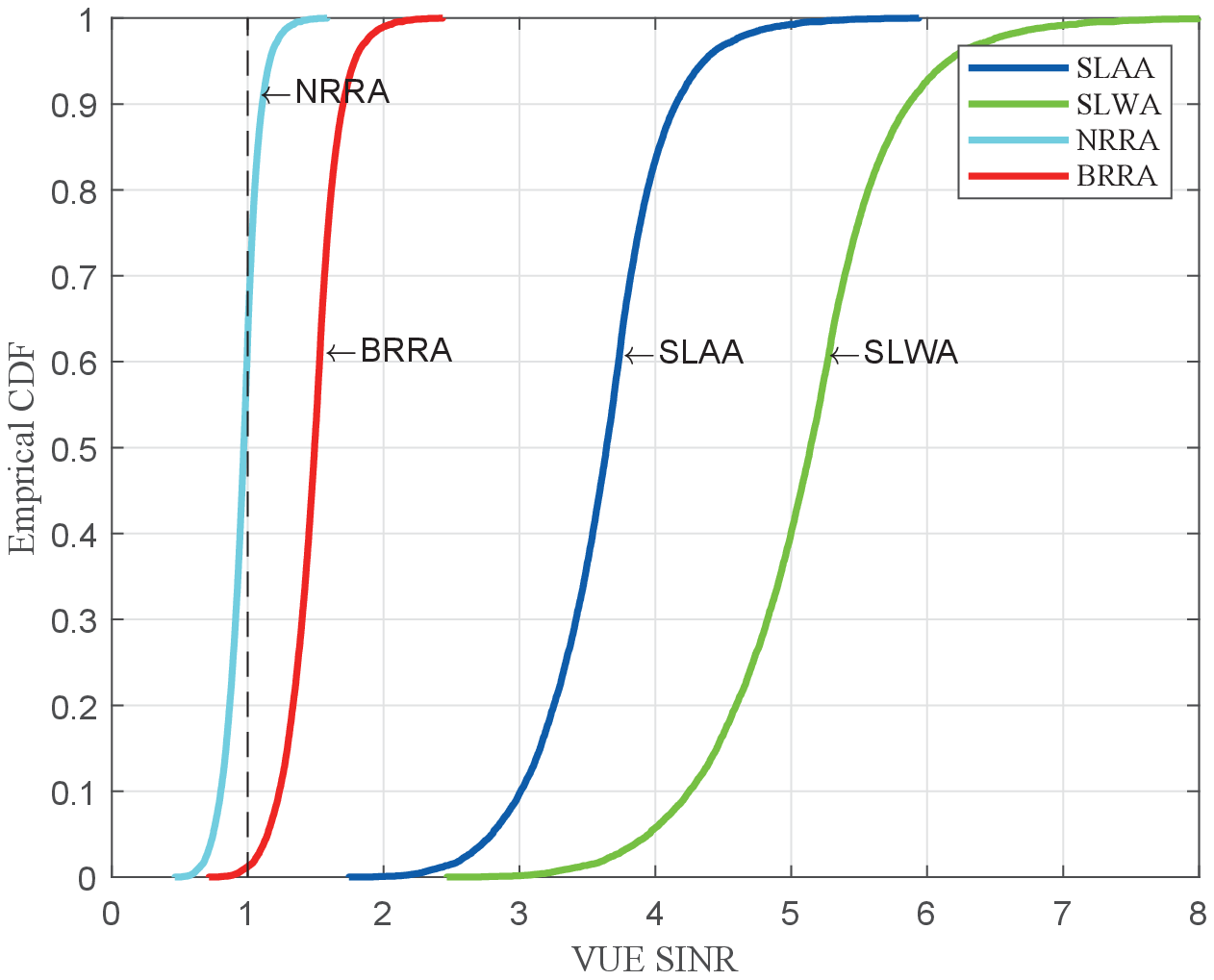}
\end{minipage}%
}%

\centering
\label{fig4}\caption{(a) The convergence of CUE transmit power versus iteration number, (b) Empirical CDF under VUE SINR.}
\vspace{-2em}
\end{figure}

In Fig. 5(a), we research the convergence of the CUE transmit power versus the iteration number.
We find that the CUE transmit power of both BRRA and NRRA converges in approximately 8 iterations. Moreover, their CUE transmit power obviously fluctuates in the process of iterations because of the requirements for the termination conditions of Bisection search method.
On the contrary, when the CUE transmit power of both SLAA and SWAA converges, their iteration number is only 2 which testifies the effectiveness of Self-learning Robust Resource Allocation approach. The reason is that the optimal power allocation can be directly obtained when the initial feasible solutions are given based on the conclusion of Section V.A.
Therefore, we conclude that the complexity of both SLAA and SWAA is lower than the complexity of both BRRA and NRRA.
We observe that the CUE transmit power of most approaches is close to $1\,w$ in Fig. 5(a). This is because they stop iterating when the optimal power allocation satisfies $p_j^{c,*}=p_{max}^c$ or $p_s^{d,*}=p_{max}^d$.
In Fig. 5(b), the empirical cumulative distribution function (CDF) of VUE SINR is utilized to evaluate the stability of the above four approaches. Because the value of $\Gamma _{\min }^d$ is 1, from the figure above, it is not difficult to understand that the outage probabilities are determined by the dotted line whose abscissa value is 1. From Fig. 5(b), the outage probability of NRRA is in a range of 0.35 to 0.45. This result shows that if the error of CSI is ignored, the network will obtain terrible security so that the vehicles cannot establish stable communication links. Thus, we propose some robust approaches with protection procedures such as SLWA, SLAA and BRRA to reduce the outage probability.
In Fig. 5(b), BRRA ensures that the outage probability is under 0.05, and both SLAA and SLWA are close to 0. However, there is a significant difference between BRRA and the two methods of Self-learning Robust Resource Allocation.
The robustness of SLAA and SLWA is overly conservative so their CDF curves are far from $\Gamma _{\min }^d$. The reason is that the affine set can contain almost all feasible solutions under the sample set and that the closed-form solutions of Self-learning Robust Resource Allocation approach almost satisfy the V2V chance constraints under the test set. On the contrary, to obtain better optimal solutions, BRRA can guarantee the $1-\beta$ percentage of solutions to satisfy the V2V chance constraints under the test set. In addition, due to the accuracy of Bisction search, Bisction search algorithm of BRRA can search for a few solutions which are beyond feasibility boundaries slightly so that they cannot satisfy the V2V chance constraints under the test set.

\begin{figure}
\centering
\subfigbottomskip=2pt
\subfigure[]
{
\begin{minipage}[c]{1.0\linewidth}
\centering
\includegraphics[scale = 0.4]{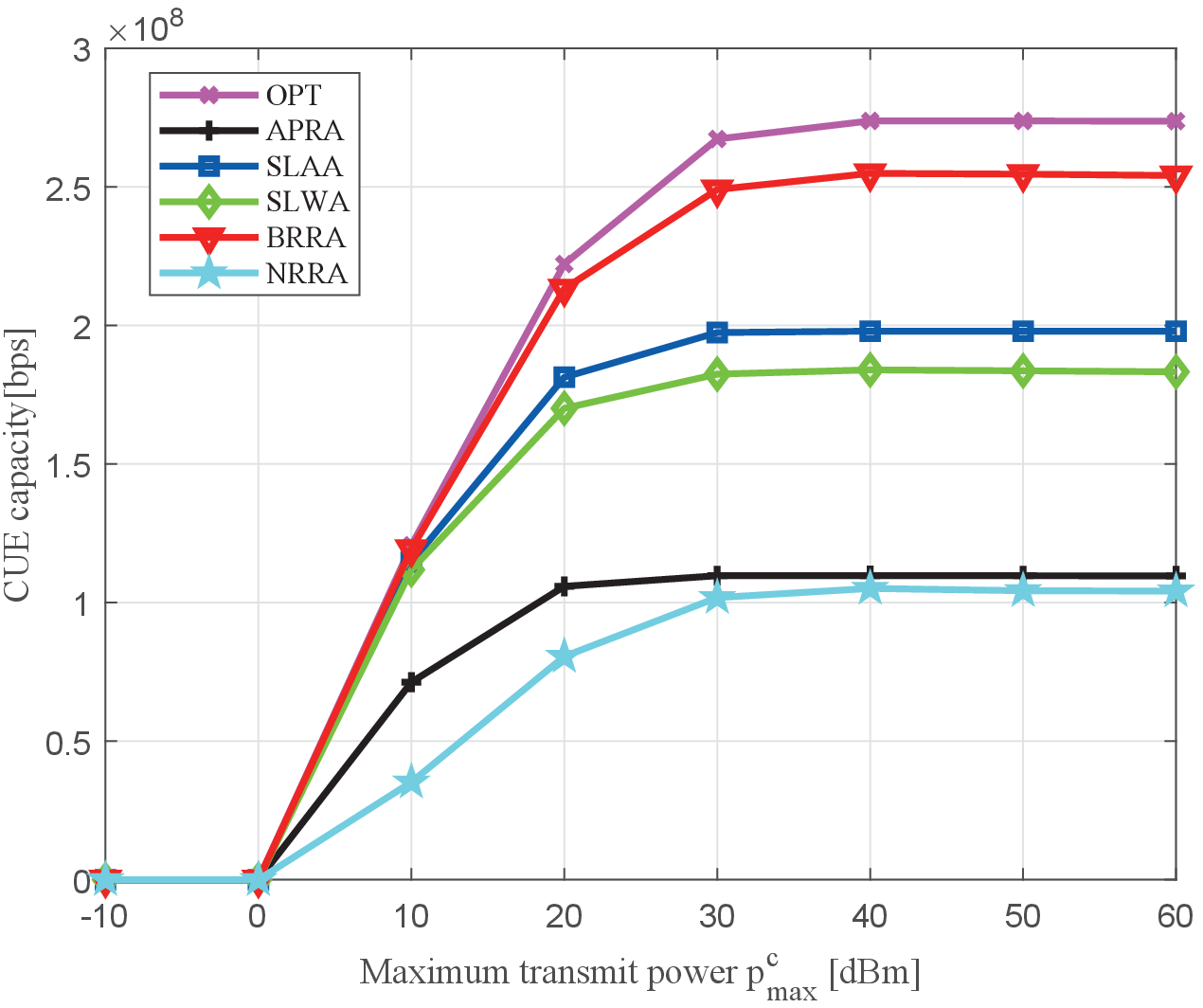}
\end{minipage}%
}
\subfigure[]
{
\begin{minipage}[c]{1.0\linewidth}
\centering
\includegraphics[scale = 0.4]{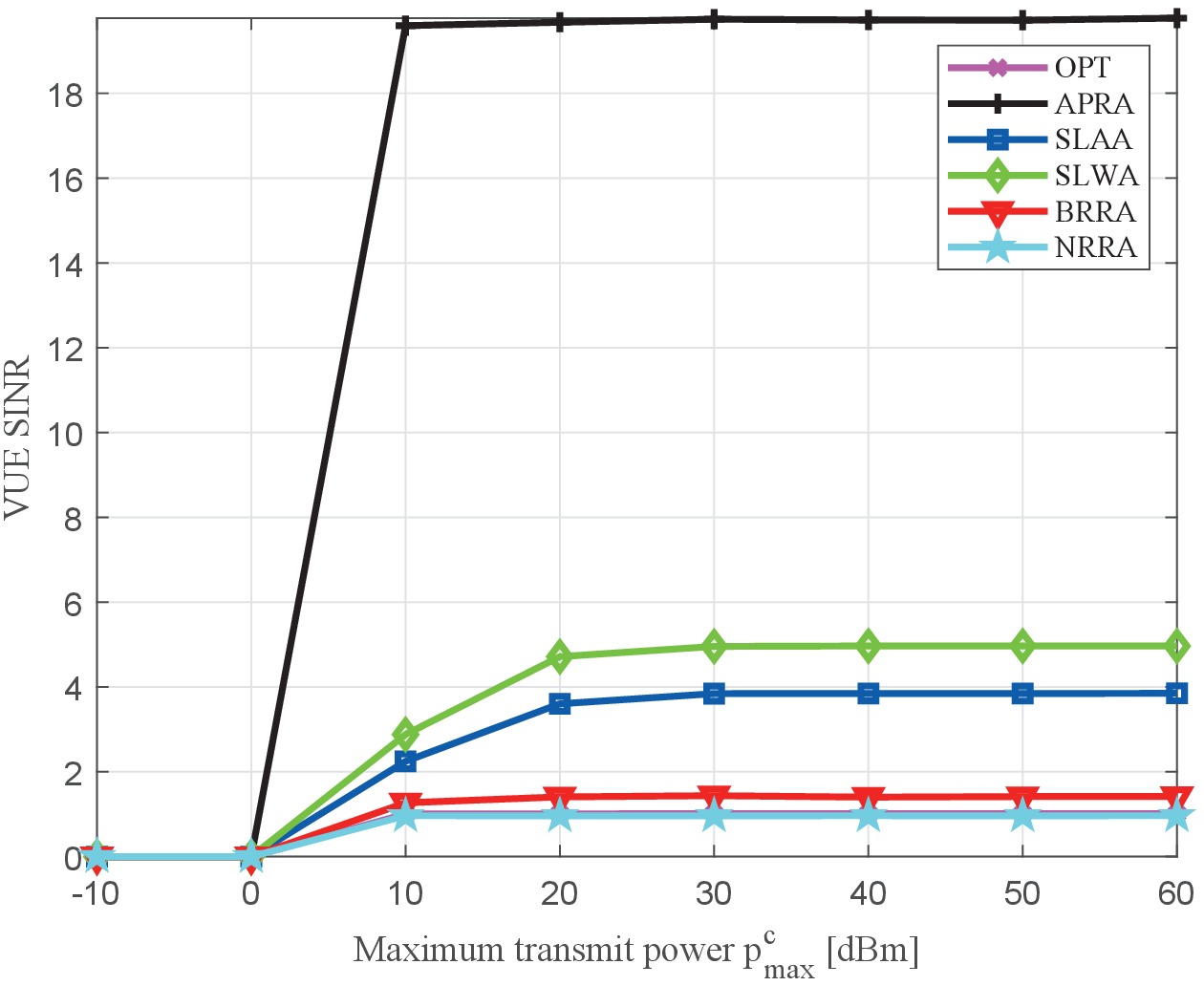}
\end{minipage}%
}%

\centering
\label{fig5}\caption{(a) CUE capacity versus the maximum transmit power of V2I communication, (b) VUE SINR versus the maximum transmit power of V2I communication.}
\vspace{-2em}
\end{figure}

Fig. 6 investigates CUE capacity and VUE SINR versus the maximum transmit power of V2I communication.
The robust resource allocation approach in \cite{8345717} is considered as a baseline method, which is represented as APRA.
Moreover, in order to verify the analyses of suboptimality in Section VI, the allocation approach of problem (37) is represented as OPT.
It is not difficult to find that the curves of both CUE capacity and VUE SINR are infinitely close to 0 when the abscissa value is smaller than $ 0\,{\rm{dBm}}$. The reason is that the requirement of the SINR threshold of V2I communication cannot be satisfied due to the small CUE transmit power.
With the increase of the abscissa value, the curves of both CUE capacity and VUE SINR increase until they reach their maximums. Then, when the abscissa value exceeds $30\,{\rm{dBm}}$, the curves of both CUE capacity and VUE SINR become steady.
The reason is that larger CUE transmit power leads to more interference of V2V communication. To decrease the interference of V2V communication, more VUE transmit power is allocated until it reaches the maximum transmit power $p_{max}^d$. Therefore, the system cannot continue allocating more transmit power $p_j^c$ to them, otherwise, the V2V QoS constraint can be violated.
Hence, the curves of both CUE capacity and VUE SINR begin to remain stable.
In Fig. 6(a), CUE capacity of OPT is the largest among all the approaches. It is not difficult to find that CUE capacity of BRRA decreases by $7\%$, CUE capacity of SLAA decreases by $27.7\%$ and CUE capacity of SLWA decreases by $32.9\%$ because the robust resource allocation approaches for dealing with the channel uncertainty are very conservative, which at many instances may not be necessary and may lead to an inefficient utilization of resources.
Furthermore, CUE capacity of BRRA is larger than the ones of SLWA, SLAA and APRA.
Because the optimal solutions of SLWA, SLAA and APRA are more conservative than BRRA, the interference of V2I communication of BRRA is smaller than others and CUE capacity of BRRA is larger.
The excessive conservatism also causes that VUE SINR of SLWA, SLAA and APRA is larger than BRRA as shown in Fig. 6(b). In Fig. 6(a), CUE capacity of SLAA is better than SLWA because SLWA needs more VUE transmit power to satisfy the SINR threshold of V2V communication. However, more VUE transmit power can cause more interference of V2I communication which can decrease CUE capacity of SLWA.
Fig. 6(a) reveals that VUE SINR of APRA is much larger than that of other approaches.
The reason is that the chance constraint in V2V links is transformed into a linear constraint by finding an approximate upper bound, which makes the SINR threshold of V2V communication much larger. Larger VUE SINR of APRA is required to satisfy larger SINR threshold and more VUE transmit power causes more interference of V2I communication, which decreases the CUE capacity of APRA. For example, when $\beta =0.05$ and ${\Gamma _{\min }^d}=1$ is given, the transformed SINR threshold of V2V communication ${\bar \Gamma _0}$ is about 19.5 according to ${\bar \Gamma _0} = \frac{{\Gamma _{\min }^d}}{{ - \ln \left( {1 - \beta } \right)}}$ \cite{8345717}. This conclusion is consistent with Fig. 6(b).
Furthermore, CUE capacity of NRRA is the smallest and VUE SINR of NRRA is less than 1 because some solutions of NRRA violate the SINR threshold of V2V communication in Fig. 6(b).

\begin{figure}
\centering
\subfigbottomskip=-2pt
\subfigure[ ]
{
\begin{minipage}[c]{1.0\linewidth}
\centering
\includegraphics[scale = 0.4]{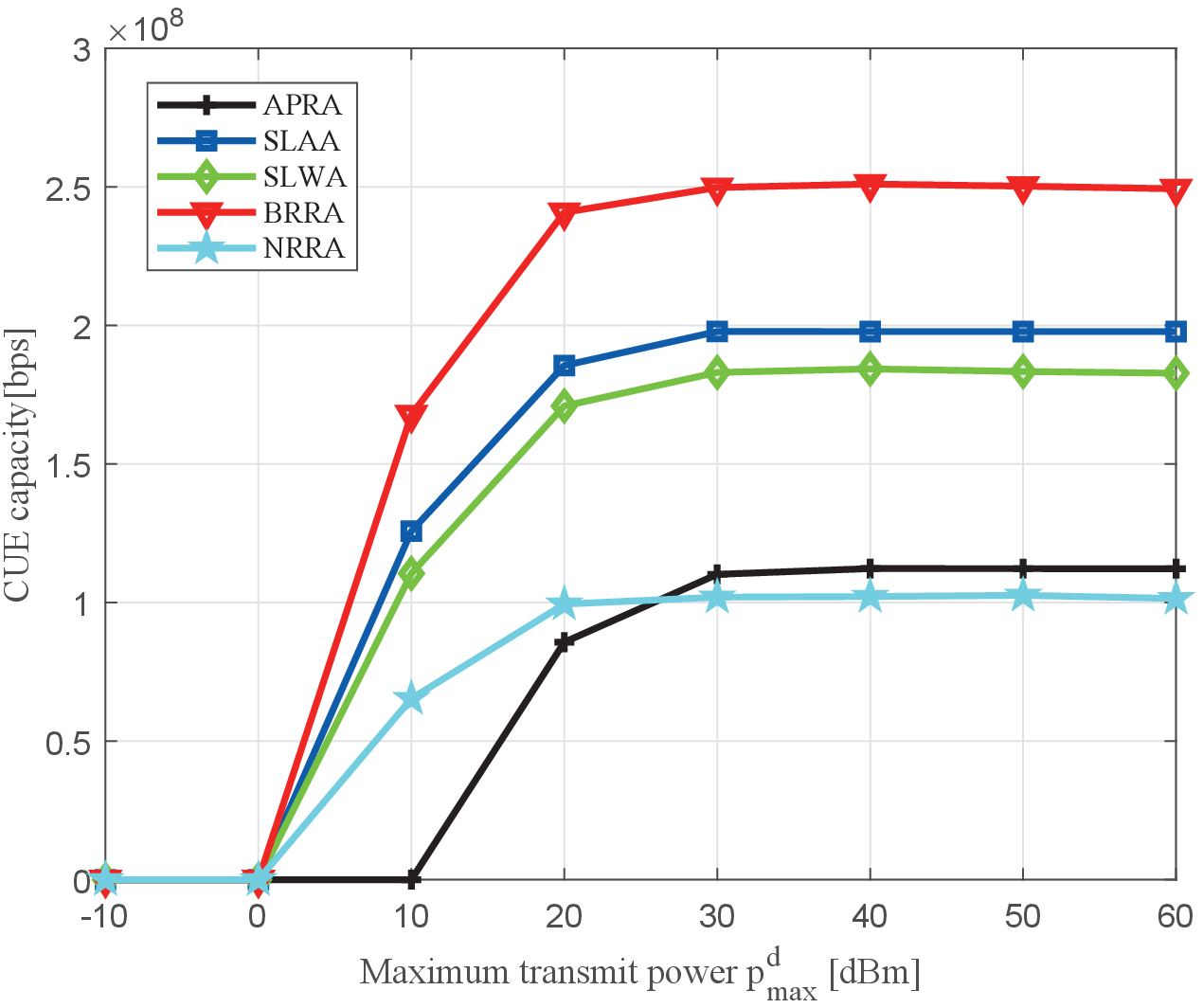}
\end{minipage}%
}\\ 
\subfigure[ ]
{
\begin{minipage}[c]{1.0\linewidth}
\centering
\includegraphics[scale = 0.4]{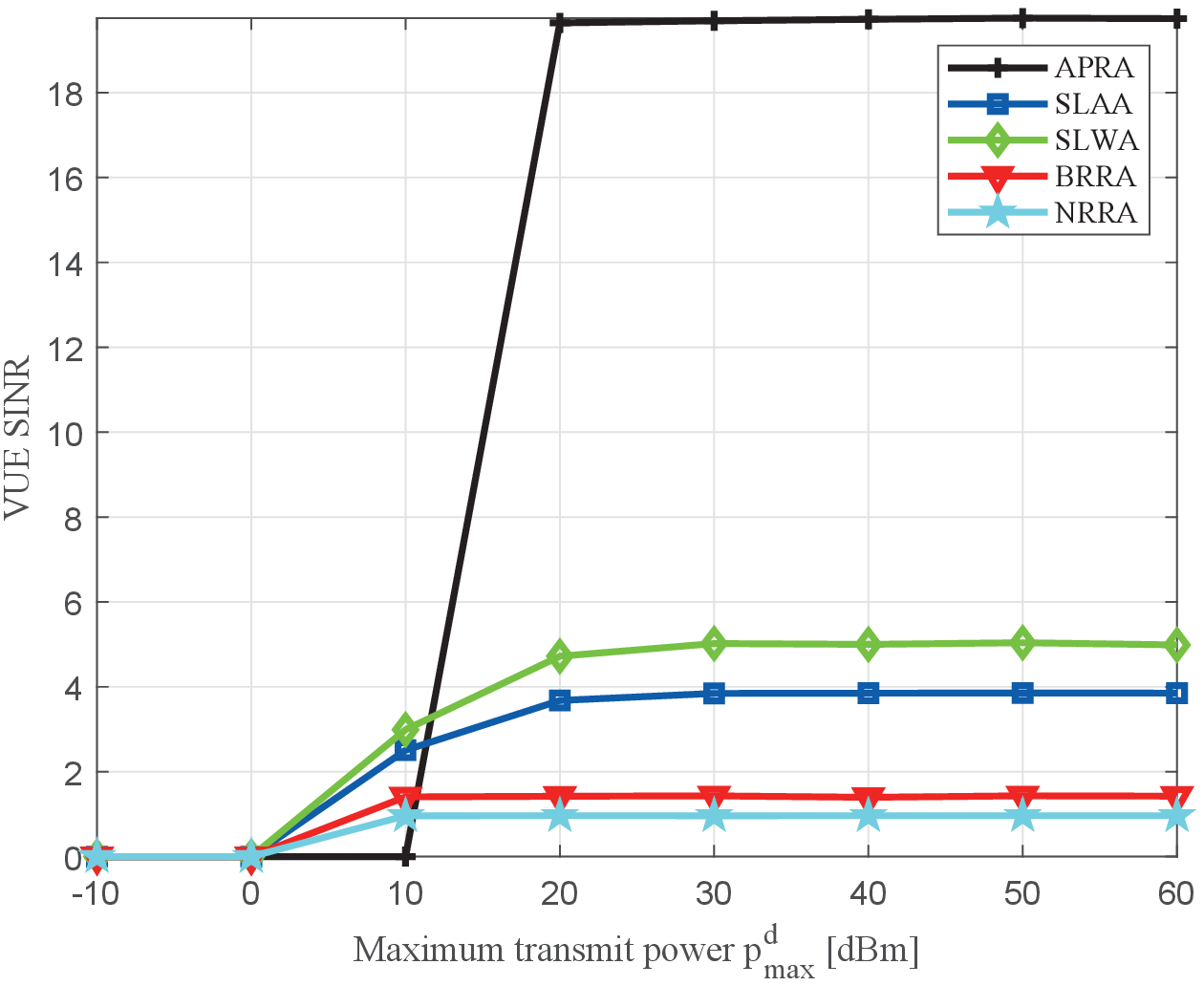}
\end{minipage}%
}%

\centering
\label{fig6}\caption{(a) CUE capacity versus the maximum transmit power of V2V communication, (b) VUE SINR versus the maximum transmit power of V2V communication.}
\vspace{-2em}
\end{figure}

Fig. 7 shows CUE capacity and VUE SINR versus the maximum transmit power of V2V communication.
In Fig. 7, the curves of both CUE capacity and VUR SINR are infinitely close to 0 when the abscissa value is close to $0\,{\rm{dBm}}$. The reason is that the SINR threshold of V2V communication is very difficult to be satisfied when the VUE transmit power is too small.
With the increase of the abscissa value, the curves of CUE capacity and VUE SINR begin to increase.
When the abscissa value is larger than $10\,{\rm{dBm}}$, the curve of APRA just begins to rise.
This is because APRA is more difficult than SLAA, SLWA and BRRA to find the feasible solution when the parameter is set in the extreme condition.
When the abscissa value exceeds $30\,{\rm{dBm}}$, their values reach the maximums and tend to be stable. This is because larger VUE transmit power results in more interference of V2I communication.
To decrease the interference of V2I communication, more CUE transmit power is allocated until it reaches the maximum transmit power $p_{max}^c$. Therefore, the system cannot continue allocating more transmit power $p_s^d$ to them, otherwise, more VUE transmit power can lead to the violation of V2I QoS constraint and smaller CUE capacity, which is contrary to our expectation.
In addition, from Fig. 7, we observe that CUE capacity of BRRA is larger than both SLWA, SLAA and APRA. The reasons are same as the analyses in Fig. 6.

\begin{figure}
\centering
\subfigbottomskip=-2pt
\subfigure[ ]{
\begin{minipage}[c]{1.0\linewidth}
\centering
\includegraphics[scale = 0.4]{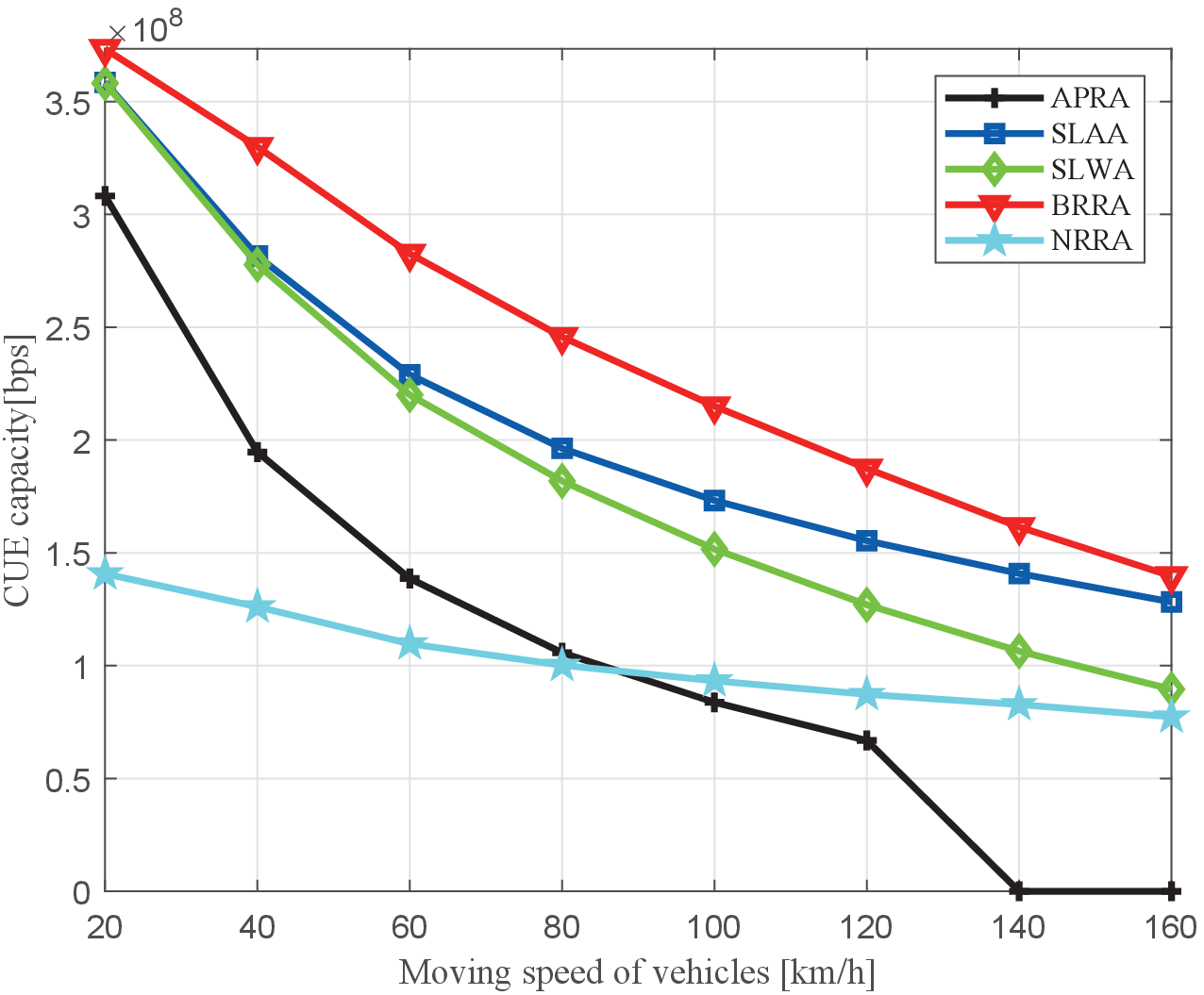}
\end{minipage}%
}\\
\subfigure[ ]{
\begin{minipage}[c]{1.0\linewidth}
\centering
\includegraphics[scale = 0.4]{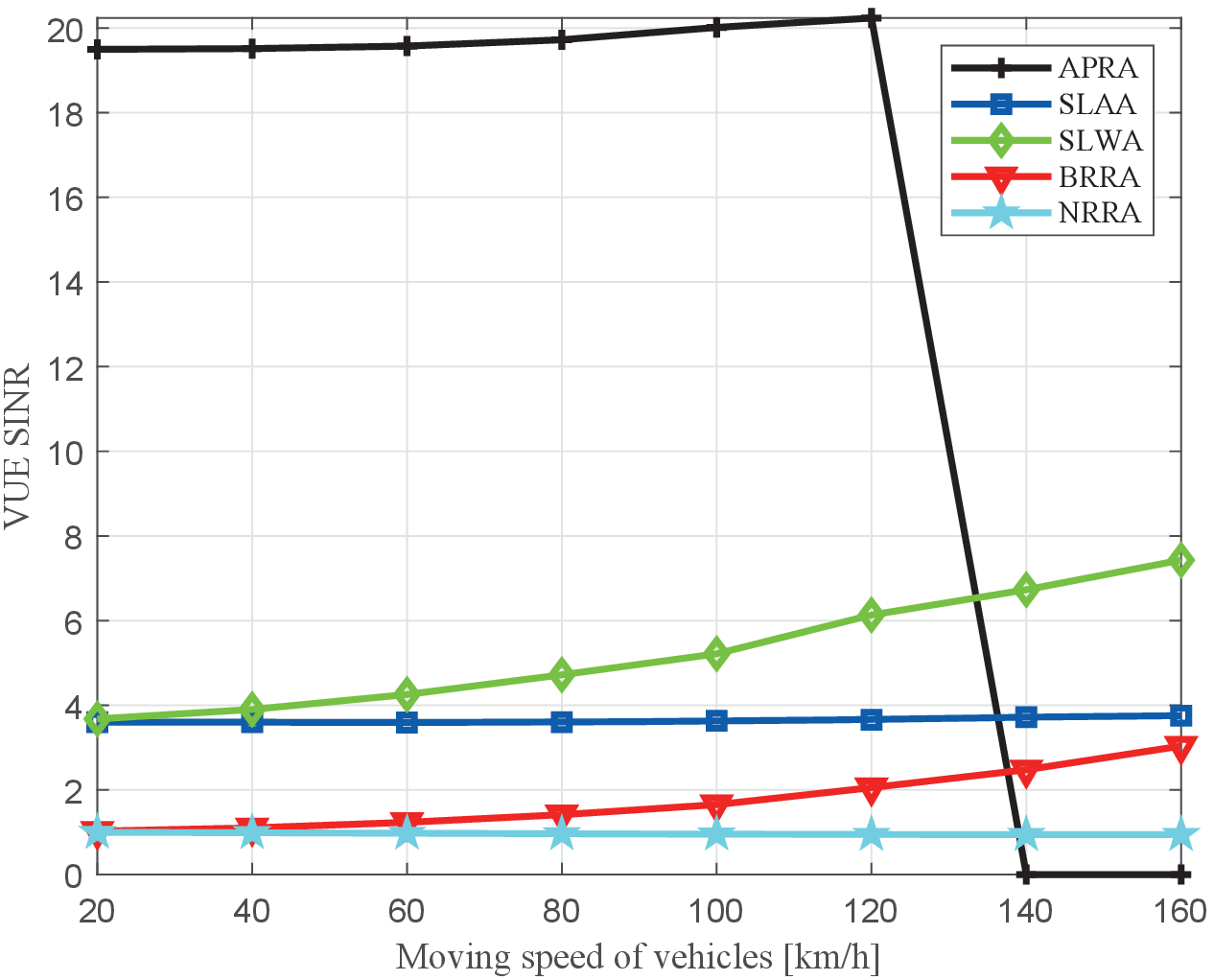}
\end{minipage}%
}%

\centering
\label{fig7}\caption{(a) CUE capacity versus moving speed of vehicles, (b) VUE SINR versus moving speed of vehicles.}
\vspace{-2em}
\end{figure}

Fig. 8 shows the curves of CUE capacity and VUE SINR under variation of the moving speed of vehicles. We evaluate the effect of Doppler shift by means of changing the moving speed of vehicles.
As shown in Fig. 8(b), the curves of SLAA, SLWA, BRRA and APRA increase with the increase of the moving speed of vehicles. The reason is that serious Doppler shift is generated and causes the increase of the uncertainty of V2V communication.
SLAA, SLWA, BRRA and APRA have to increase the transmit power of V2V communication to reduce the uncertainty so their VUE SINR increases. Moreover, the curve of NRRA remains steady since NRRA neglects Doppler shift, which only generates the influence in the small-scale fading.
From the curves in Fig. 8(a), we observe that the curves of SLAA, SLWA, BRRA and APRA decrease with the increase of the moving speed of vehicles. This is because more transmit power of V2V communication is allocated, which causes more interference of V2I communication.
When the speed of vehicles is larger than $140\;km/h$, both CUE capacity and VUE SINR of APRA drop to 0. The reason is that the QoS constraint of V2I communication cannot be satisfied with the increase of the transmit power of V2V communication.
Thus, besides NRRA, all of CUE capacity decreases. In addition, we observe that CUE capacity of BRRA is the biggest of all.

\begin{figure}
\centering
\subfigbottomskip=-3pt
\subfigure[ ]{
\begin{minipage}[c]{1.0\linewidth}
\centering
\includegraphics[scale = 0.4]{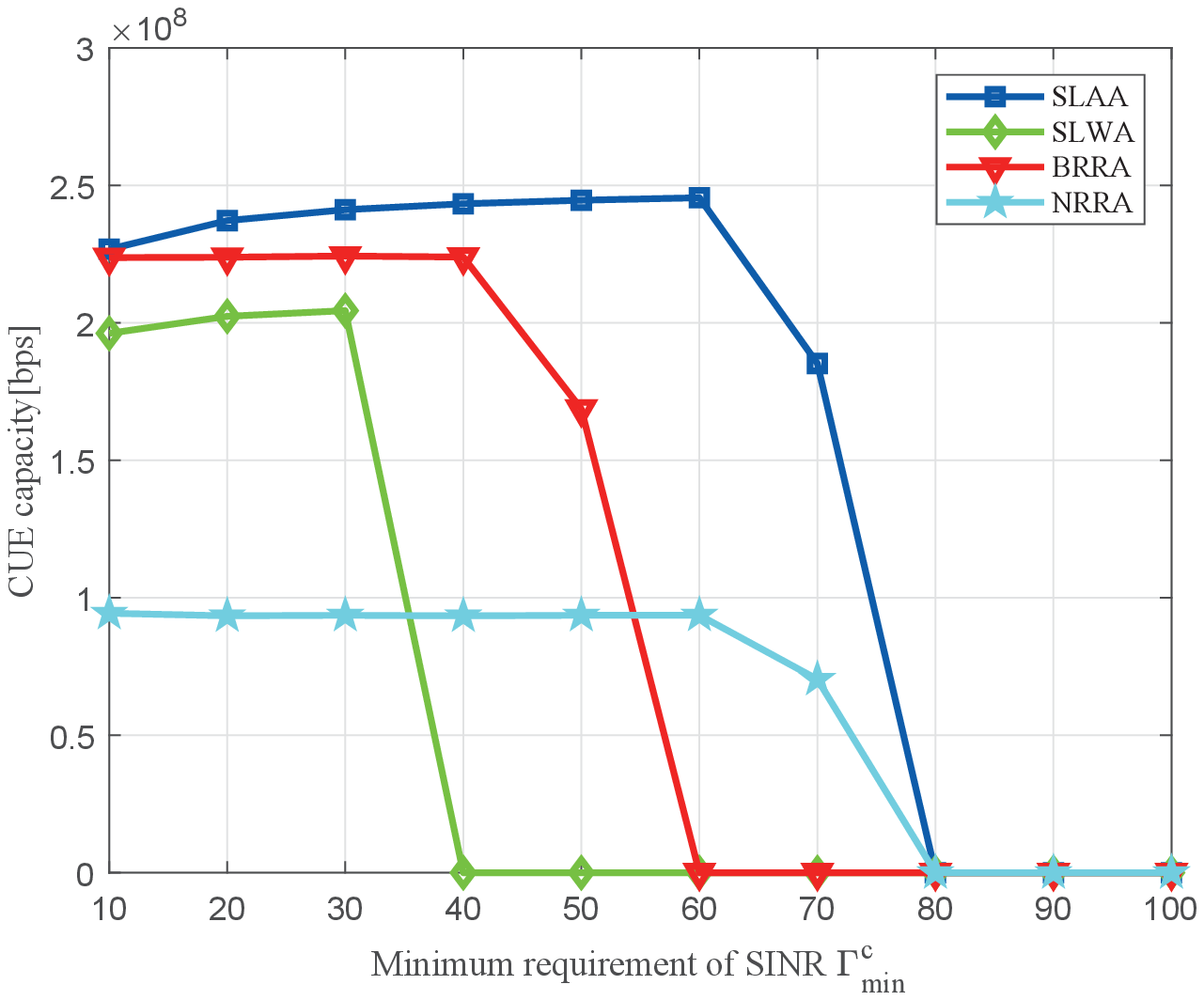}
\end{minipage}%
}\\
\subfigure[ ]{
\begin{minipage}[c]{1.0\linewidth}
\centering
\includegraphics[scale = 0.4]{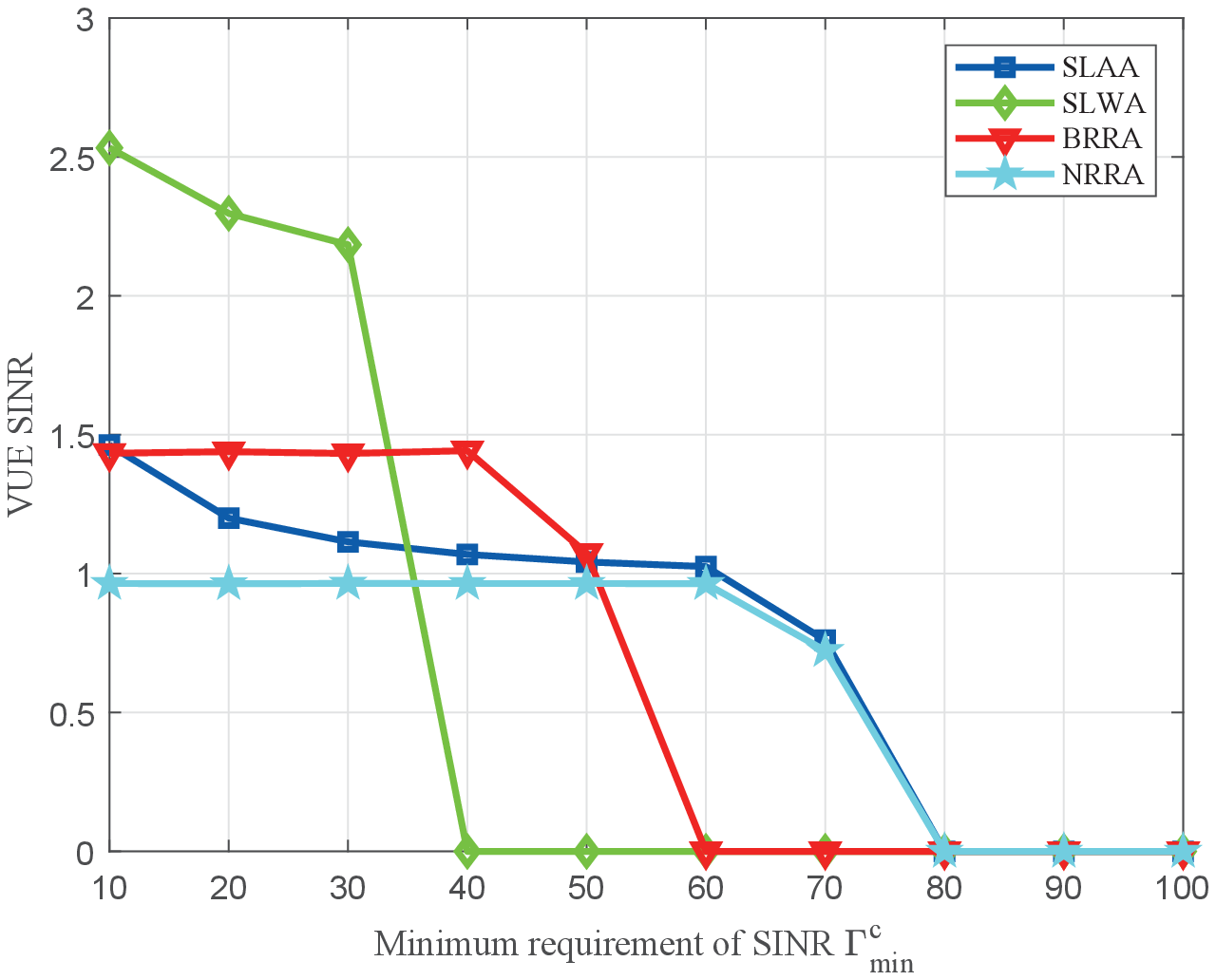}
\end{minipage}%
}%

\centering
\label{fig8}\caption{(a) CUE capacity versus the SINR threshold of V2I communication, (b) VUE SINR versus the SINR threshold of V2I communication.}
\vspace{-2em}
\end{figure}

Fig. 9 researches CUE capacity and VUE SINR under variation of the SINR threshold of V2I communication.
From the curves of Fig. 9(a), the curves of both SLAA and SLWA slightly rise and the curves of both BRRA and NRRA remain stable before the abscissa value increases to 30. The reason why the curves of SLAA and SLWA slightly rise is that the increase of the SINR threshold of V2I communication can influence the value of the initial feasible solution $\boldsymbol{\tilde p}_s^d$ which also influence VUE SINR of SLAA and SLWA as shown in Fig. 9(b). Then, when the abscissa value exceeds 30, the curves of these methods start to decrease in succession. We observe the similar tendency in Fig. 9(b). The reason is that the SINR threshold cannot be achieved due to the limited maximum transmit power of V2V communication. Then, the feasible probability of the problem decreases so that CUE capacity also decreases. Compared with other approaches, the curves of SLWA are the earliest to decrease because SLWA needs more VUE SINR. More VUE SINR can result in larger VUE transmit power which bring larger interference and smaller capacity of V2I communication.

\begin{figure}
\centering
\subfigbottomskip=-2pt
\subfigure[ ]{
\begin{minipage}[c]{1.0\linewidth}
\centering
\includegraphics[scale = 0.4]{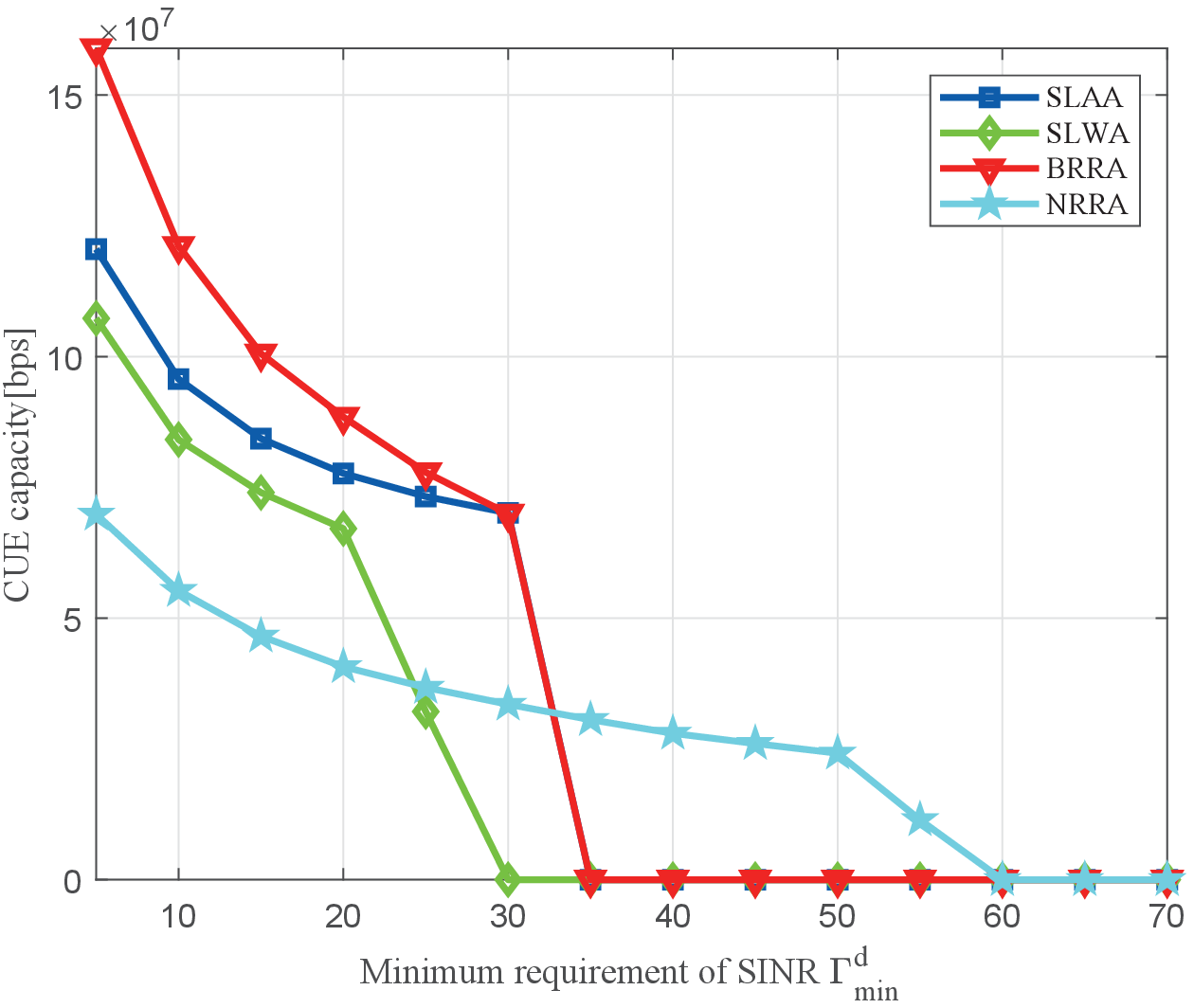}
\end{minipage}%
}\\
\subfigure[ ]{
\begin{minipage}[c]{1.0\linewidth}
\centering
\includegraphics[scale = 0.4]{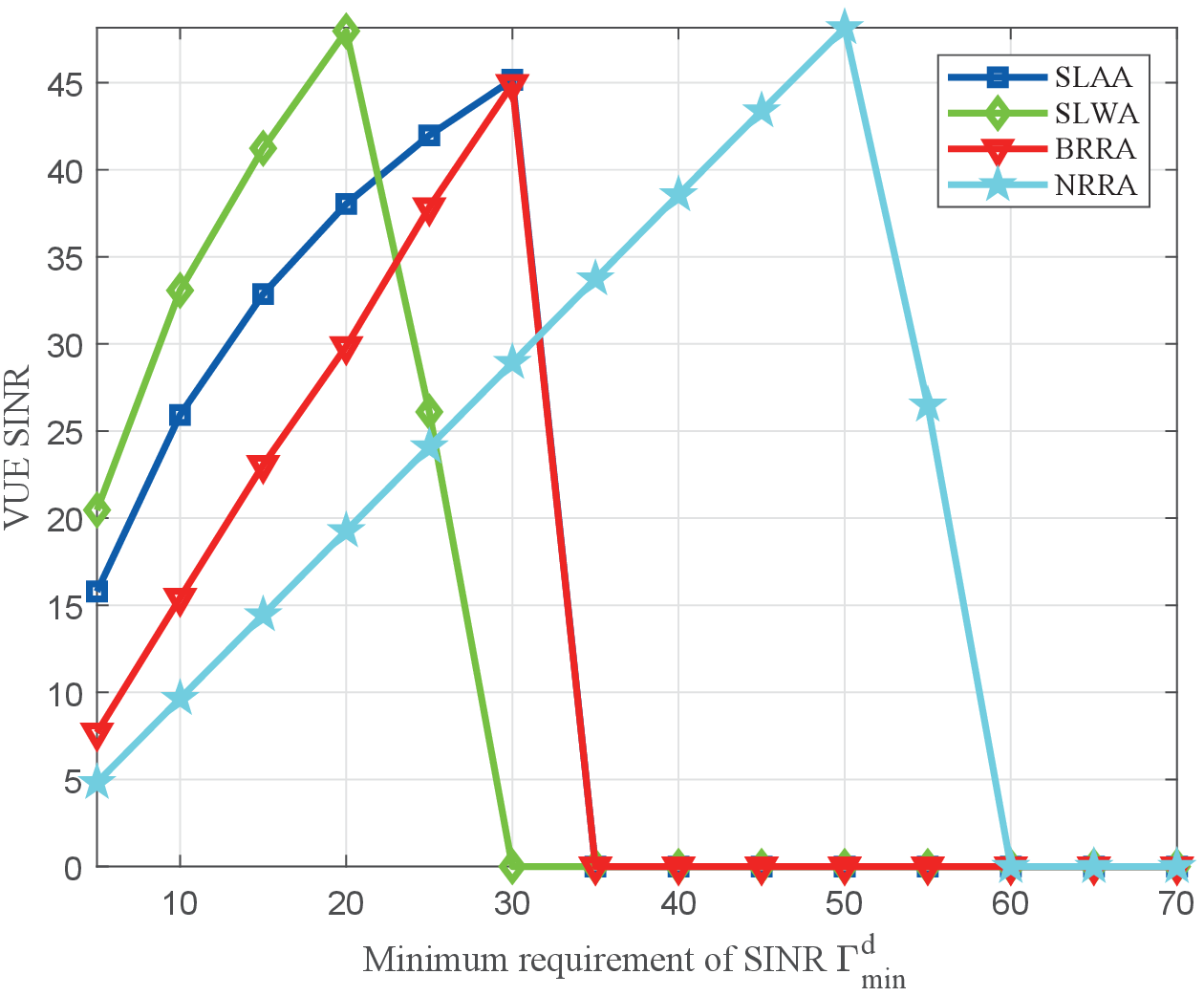}
\end{minipage}%
}\\
\subfigure[ ]{
\begin{minipage}[c]{1.0\linewidth}
\centering
\includegraphics[scale = 0.4]{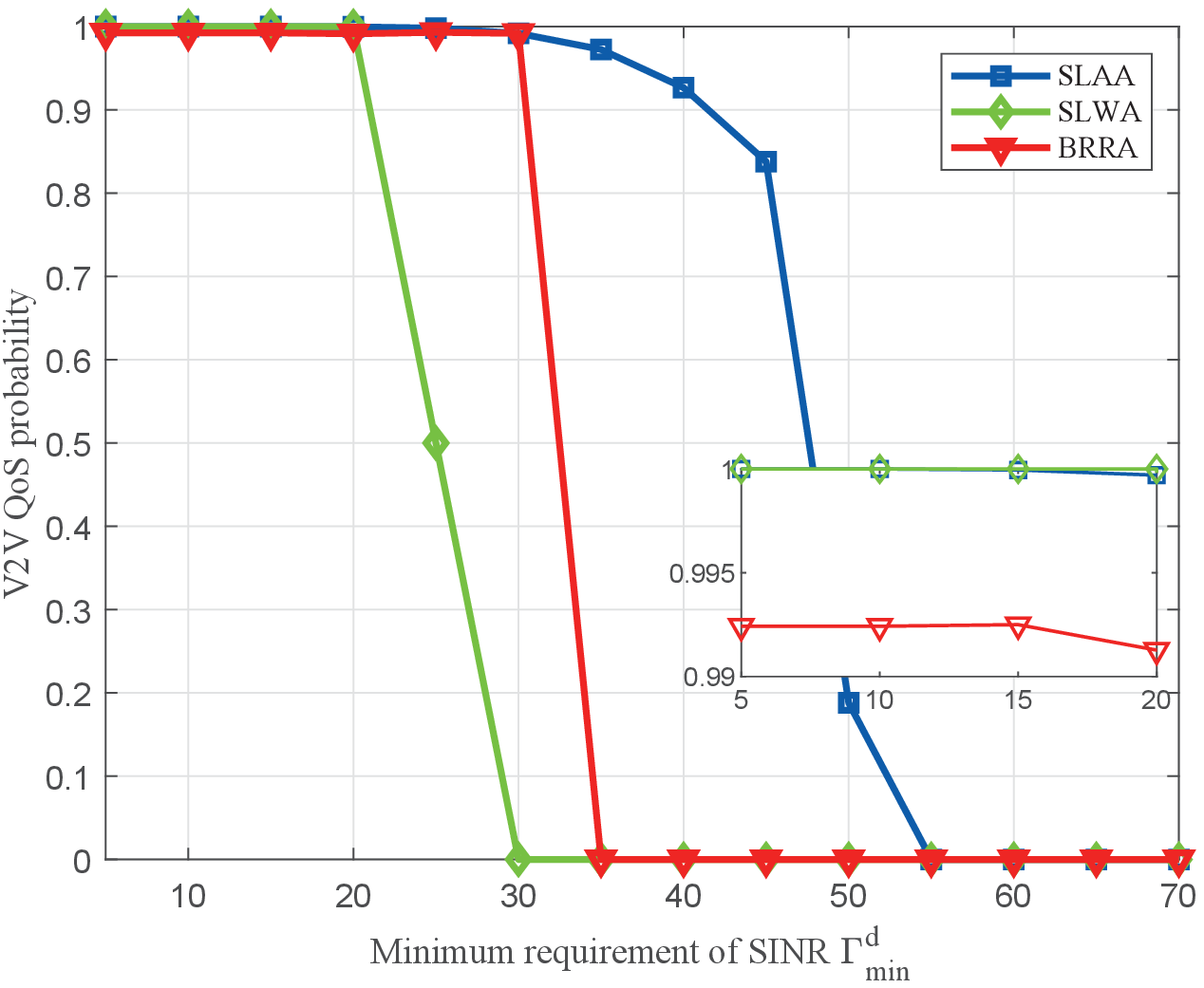}
\end{minipage}%
}

\centering
\label{fig9}\caption{(a) CUE capacity versus the SINR threshold of V2V communication, (b) VUE SINR versus the SINR threshold of V2V communication, (c) V2V QoS probability versus the SINR threshold of V2V communication.}
\vspace{-2em}
\end{figure}

Fig. 10 shows the curves of CUE capacity, VUE SINR and V2V QoS probability under variation of the SINR threshold of V2V communication.
In order to better meet the high reliability of V2X communication, the outage probability is set as $\beta =0.01$.
From the curves in Fig. 10(a) and 10(b), when the abscissa value is in the range of 0 to 20, the curves of CUE capacity decrease while the curves of VUE SINR increase. The reason is that the feasible solutions can be obtained in this interval so that more transmit power of V2V communication is required for satisfying the SINR threshold of V2V communication.
BRRA and SLAA can obtain larger capacity than others because they consume less VUE transmit power to reach the SINR threshold of V2V communication and this causes less interference of V2I communication.
In the larger version of Fig. 10(c), we do not consider the non-robust case because the non-robust case is not limited by the outage probability.
It can be clearly seen that V2V QoS probability of SLAA, SLWA and BRRA is in the range of 0.99 to 1. It demonstrates that the outage probability of V2V communication is guaranteed when the abscissa value is less than 20.
Then, when the abscissa value is greater than 20, the curves of all the approaches begin to sharply decrease in succession.
This is because the VUE transmit power cannot continuously increase due to the limited maximum transmit power of V2V communication. The result is that V2V QoS probability decreases and the outage probability of V2V communication can not be guaranteed. Thus, the curves of Fig. 10(a) and 10(b) descend to 0.
Moreover, the curve of SLWA is the earliest to drop to $0$ because the VUE SINR of SLWA is the largest and its conservatism is also the highest. Higher conservatism requires larger transmit power requirements. Thus, when the SINR threshold of V2V communication becomes bigger, transmit power requirements can no longer be satisfied and V2V QoS probability can also not be guaranteed.

In summary, as shown in Fig. 6(a)-10(a), CUE capacity of BRRA is the largest and the one of Non-robust is the smallest among all of the approaches. CUE capacity of SLAA is larger than SLWA. According to the analysis of Sections IV and V, we can draw the conclusion that the complexity of SLAA and SLWA is lower than that of BRRA. Thus, further analysis shows that there exists a trade-off between performance and computation complexity for Self-learning Robust Resource Allocation and BARR. We can also draw the conclusion that VUE SINR of BRRA is lower than the ones of SLAA and SLWA.
Therefore, there exists a trade-off between robustness and CUE capacity for Self-learning Robust Resource Allocation and BRRA.
In addition, we discover that the performance can be transformed when we change the relevant parameters in above experiments. The results show that the proposed methods may achieve the best performance when they are in suitable configurations.

\section{Conclusions}

In this article, we investigated the joint spectrum and power allocation problem for V2X communication with imperfect CSI.
In the realistic scenario, the gNB can acquire the accurate CSI of V2I communication and the estimated CSI with error of V2V communication.
Then, two different robust approaches were designed to solve the above problem. One was Bernstein Approximation-based Robust Resource Allocation approach.
Then, to further reduce computational complexity, the other was Self-learning Robust Resource Allocation approach, which included a learning method and an analytical mapping method.
Then, we illustrated that the above approaches can significantly improve the CUE capacity.
Finally, in subsequent research, we attended to investigate a more complex resource problem approach where each CUE can share the spectrum resource with multiple VUEs and each VUE can reuse the resource of multiple CUEs.

\footnotesize
\bibliographystyle{IEEEtran}
\bibliography{mybibfile}

\end{document}